\documentclass[pdflatex,sn-apa, iicol]{sn-jnl}

\usepackage{graphicx}%
\usepackage{multirow}%
\usepackage{amsmath,amssymb,amsfonts}%
\usepackage{amsthm}%
\usepackage{mathrsfs}%
\usepackage[title]{appendix}%
\usepackage{xcolor}%
\usepackage{textcomp}%
\usepackage{manyfoot}%
\usepackage{booktabs}%
\usepackage{algorithm}%
\usepackage{algorithmicx}%
\usepackage{algpseudocode}%
\usepackage{listings}%
\usepackage{bbm}
\usepackage{mathtools}

\usepackage{caption}

\theoremstyle{thmstyleone}%
\theoremstyle{thmstyletwo}%
\newtheorem{remark}{Remark}%
\theoremstyle{thmstylethree}%
\begin{document}

\title[1]{\centering On the Localization of Checkerboarding in Multiaxial Stress Regions under SIMP Penalization} 

\author*[1]{\fnm{Iulian} \sur{Paunel}}\email{iulian.paunel@tu-darmstadt.de}

\author[1]{\fnm{Jonathan} \sur{Stollberg}}\email{jonathan.stollberg@tu-darmstadt.de}

\author[1]{\fnm{Dominik} \sur{Schillinger}}\email{dominik.schillinger@tu-darmstadt.de}

\affil[1]{\orgdiv{Institute for Mechanics, Computational Mechanics Group}, \orgname{Technical University of Darmstadt}, \country{Germany}} 

\abstract{

Checkerboard patterns are a well-known numerical artifact in density-based topology optimization using the Solid Isotropic Material with Penalization (SIMP) method and linear finite elements. Existing explanations based on mixed-field incompatibility or locking-induced stiffness overestimation explain the artificial stiffness of checkerboard layouts but do not clarify their characteristic spatial localization. In this work, we show that checkerboard patterns systematically emerge in multiaxial load-transfer regions, whereas predominantly uniaxial stress regions remain checkerboard-free. Through systematic numerical investigations, we demonstrate that checkerboarding originates where continuous intermediate densities are mechanically favorable for multiaxial load transfer but are suppressed by SIMP penalization. Due to the characteristic behavior of linear elements, checkerboard layouts provide an artificially stiff discrete substitute for these penalized intermediate-density regions. In contrast, uniaxial load paths naturally favor continuous solid struts, rendering checkerboards mechanically disadvantageous. Our findings provide a unified mechanical interpretation of checkerboarding as the interplay between global stress states, SIMP penalization, and element-level locking, thereby explaining both its origin and the spatial localization.}


\maketitle
\section{Introduction}\label{sec1}

In the standard framework of density-based topology optimization, structural design problems are often solved using the Optimality Criteria Method (OCM) established by \citet{bendsoe1988generating}, frequently in combination with the Solid Isotropic Material with Penalization (SIMP) approach subsequently introduced by \citet{bendsoe1989optimal}. However, such formulations are notoriously affected by numerical artifacts, including non-uniqueness of the solution, mesh dependency, and checkerboard patterns \citep{sigmund1998numerical}.

The increasing integration of topology optimization with Additive Manufacturing (AM) has significantly intensified the need for a rigorous elimination of numerical artifacts and intermediate density regions, Since AM requires strictly binary, manufacturable layouts without ambiguity \citep{brackett2011topology, liu2018current}.
The growing integration of Artificial Intelligence (AI) and Machine Learning (ML) based techniques into topology optimization further increases the importance of understanding numerical artifacts. For example, as neural networks are increasingly used to accelerate optimization procedures \citep{banga20183d, sosnovik2019neural, stollberg2025multiscale}, unresolved numerical inconsistencies may become implicitly embedded into data-driven models. 

To address this issue, numerous regularization techniques have been developed over the last three decades. These include sensitivity filtering \citep{sigmund1994design, sigmund1998numerical, zhou2001checkerboard}, perimeter constraints \citep{haber1996new, ambrosio1993optimal, petersson1999some}, gradient constraints \citep{petersson1998slope, borrvall2001topology}, and local patch restrictions that actively suppress checkerboard formation \citep{johnson1982analysis, bendsoe2004topology}. Moreover, \citet{poulsen2002simple} proposed an effective scheme to prevent the formation of artificial single-node hinges. Conceptually distinct approaches involve the use of non-conforming finite element discretizations \citep{jang2003checkerboard}, where elements are not exclusively connected via their corners, physically precluding the emergence of checkerboard patterns. Alternatively, higher-order displacement interpolation can avoid checkerboarding, albeit at a significantly increased  computational cost. Regarding element shape dependencies, \citet{talischi2009honeycomb} demonstrated that hexagonal elements form ring-like artifacts, while rectangular elements tend to produce checkerboard patterns. Notably, the majority of these regularization methods simultaneously mitigate multiple of the aforementioned issues.

This broad range of classical regularization methodologies forms the foundation for many subsequent techniques, which are predominantly extensions of these classical concepts rather than entirely new paradigms \citet{sigmund2013topology}. For instance, alternative penalization schemes such as RAMP \citep{stolpe2001alternative, stolpe2001trajectories} have been proposed. Furthermore, two-field \citep{bruns2001topology, bourdin2001filters, lazarov2011filters, kawamoto2011heaviside} and three-field \citep{guest2004achieving, sigmund2007morphology, xu2010volume} SIMP approaches were introduced as advanced filtering techniques to also eliminate grey transition boundaries. These methods involve additional operations on the density field and consequently require the application of the chain rule to calculate sensitivities. More recently, these various techniques have been rigorously unified into a generalized nonlinear filtering framework based on quasi-arithmetic means \citep{wadbro2015quasi, hagg2017nonlinear}, ensuring mathematical well-posedness and efficient sensitivity analysis. 

While filtering methods successfully suppress the visible symptoms of numerical artifacts, achieving a true physical resolution requires a deeper understanding of their underlying mechanics. To date, the origin of checkerboard patterns has primarily been interpreted from two distinct perspectives, presented by \citet{jog1996stability} and \citet{diaz1995checkerboard}. On the one hand, checkerboarding has been linked to stability issues analogous to those encountered in mixed finite element formulations in the context of the Ladyzhenskaya-Babu\v{s}ka-Brezzi (LBB) condition \citep{jog1996stability}. However, due to the bounded nature of the density variable and the use of decoupled update schemes in density-based optimization, this analogy cannot be strictly upheld, although it points to incompatibilities in the discretization. On the other hand, \citet{diaz1995checkerboard} interpreted checkerboard patterns mechanically as periodic microstructures. By evaluating its effective material properties, they demonstrated that standard four-node bilinear elements artificially overestimate the stiffness of such configurations, which is also referred to as locking. This microstructure-based viewpoint is closely related to homogenization approaches commonly used in multiscale material optimization \citep{gangwar2021concurrent}.

In practice, however, checkerboard patterns do not emerge uniformly throughout the design domain, but localize in specific regions. Neither existing explanation provides a consistent rationale for this spatial localization. In this work, we will first show that the occurrence of checkerboarding is governed by the underlying stress state: checkerboard patterns arise in regions subjected to multiaxial stress, while regions dominated by uniaxial stress remain free of such artifacts. Based on this observation, we will then provide a mechanics-based interpretation of checkerboarding as the result of an interplay between multiaxial load transfer, SIMP penalization, and the locking behavior of bilinear elements.

The remainder of this paper is structured as follows. In Section~\ref{sec2}, we introduce some theoretical and algorithmic background of density-based topology optimization. In Section~\ref{literature_chapter}, we critically review the two existing explanation approaches of checkerboarding in terms of an LBB-type instability and locking. In Section~\ref{main}, we present systematic numerical investigations that reveal the localization of checkerboard patterns and its relation to the underlying stress state.  Furthermore, we provide a mechanical interpretation of the observed behavior and discuss its implications. Finally, Section~\ref{fazit} concludes the paper.

\begin{remark}
For clarity of presentation, the developments in this work are restricted to topology optimization problems on two-dimensional domains under plane-stress conditions. All finite element discretizations are based on quadrilateral four-node bilinear ($Q_4$) and nine-node biquadratic ($Q_9$) elements. We emphasize that the all arguments in this work carry over directly to three-dimensional settings employing hexahedral finite elements.
\end{remark}

\section{Some background on density-based topology optimization using OCM and SIMP}\label{sec2}

This section outlines the theoretical foundations of density-based topology optimization that are required for the subsequent analysis. For further details, we refer to \cite{bendsoe2004topology}. 

\subsection{Problem formulation}\label{formulierung_minimierungsproblem}
Let $\Omega \subset \mathbb{R}^d$ with $d \in \left\{2, 3\right\}$ be a bounded set with volume $V$, consisting of a material domain $\Omega^{\mathrm{mat}} \subseteq \Omega$. The material is assumed to be homogeneous, isotropic, and linearly elastic, described by the stiffness tensor $E^0_{ijkl}$. For any spatial point $\mathbf{x} \in \Omega$, the indicator function $\mathbbm{1}_{\Omega^\mathrm{{mat}}}(\mathbf{x})$ is defined as:
\begin{equation}
    \mathbbm{1}_{\Omega^\mathrm{{mat}}}(\mathbf{x}) = 
    \begin{cases} 
        1 & \text{if } \mathbf{x} \in \Omega^\mathrm{{mat}}\,, \\ 
        0 & \text{otherwise} \,.
    \end{cases}
\end{equation}
Using this indicator function, the stiffness is defined as $E_{ijkl}(\mathbf{x}) = \mathbbm{1}_{\Omega^\mathrm{{mat}}}(\mathbf{x})E^0_{ijkl}$. The volume $V_\mathrm{{mat}}$ of $\Omega^\mathrm{{mat}}$ and the prescribed volume fraction $v_\mathrm{{mat}}$ are defined as:
\begin{equation} 
    V_\mathrm{{mat}} \coloneqq \int_\Omega \mathbbm{1}_{\Omega^\mathrm{{mat}}}(\mathbf{x}) \, d\Omega\,,
    \quad v_\mathrm{{mat}} \coloneqq \frac{V_\mathrm{{mat}}}{V}\,. 
\end{equation}
Discretizing the domain into $n_{\mathrm{el}}$ finite elements, the bilinear and linear forms read:
\begin{equation}
	a(\mathbf{u}, \mathbf{v}) = \sum_{e=1}^{n_{\mathrm{el}}} \mathbf{v}_e^\mathrm{T} \mathbf{K}_e \mathbf{u}_e\,, \quad
	l(\mathbf{v}) = \sum_{e=1}^{n_{\mathrm{el}}} \mathbf{v}_e^\mathrm{T} \mathbf{f}_e \,,
\end{equation}
where $\mathbf{K}_e$ is the element stiffness matrix, $\mathbf{f}_e$ is the element force vector, and $\mathbf{u}_e$ and $\mathbf{v}_e$ are the element displacement and virtual displacement vectors, respectively.

To perform the discrete 0-1 topology optimization, we introduce the discrete indicator variable $\mathbbm{1}_e \in \{0, 1\}$ for each element $e$, where $\mathbbm{1}_e = 1$ indicates a solid material element and $\mathbbm{1}_e = 0$ denotes a void element. The set of admissible global stiffness matrices $\mathbf{K}$ is obtained by gathering the active element contributions. Using the standard finite element assembly operator $\mathop{\text{A}}_{1}^{n_{\mathrm{el}}}$, which maps and sums the local element stiffness matrices into the global system matrix, the admissible set is defined as:
\begin{equation}\label{eq:K_adm}
    \begin{aligned}
        \mathcal{K}_{\mathrm{adm}} \coloneqq \Bigg\{ &\mathbf{K} = \mathop{\text{\huge A}}_{e=1}^{n_{\mathrm{el}}} \mathbbm{1}_e \mathbf{K}_e^0 \;\Bigg|\; \\ 
        &\sum_{e=1}^{n_{\mathrm{el}}} \mathbbm{1}_e V_e = V_{\mathrm{mat}} \Bigg\} \,,
    \end{aligned}
\end{equation}
where $\mathbf{K}_e^0$ denotes the stiffness matrix of a solid base element and $V_e$ the volume of element $e$.

While topology optimization frameworks can incorporate various objective functions (such as stress, displacement, or mechanism design), this work focuses on the classic minimum compliance problem, effectively maximizing global structural stiffness by minimizing external work. Subject to the equilibrium condition, the discrete optimization problem is written as:
\begin{equation}\label{eq:minimierungsproblem}
	\begin{aligned}
		\inf_{\mathbf{u}, \mathbf{K} \in \mathcal{K}_{\mathrm{adm}}} \quad & l(\mathbf{u}) \\
		\text{s.t.} \quad & a(\mathbf{u},\mathbf{v}) = l(\mathbf{v}) \quad \forall \,\mathbf{v} \,.
	\end{aligned}
\end{equation}
Due to the binary constraint imposed by $\mathbbm{1}_e$, no sensitivities or gradients can be mathematically defined. Therefore, a continuous relaxation of the problem is required.

\subsection{The SIMP method}\label{SIMP_OCM}
The discrete binary problem is converted into a continuous one using the SIMP (Solid Isotropic Material with Penalization) method \citep{bendsoe1989optimal}. Intermediate densities $\rho_e \in [\rho_{min}, 1]$ are now permitted for each element $e$. The penalized stiffness matrix is then defined as:
\begin{equation}\label{SIMP}
	\mathbf{K}_e(\rho_e) \coloneqq \rho_e^p \mathbf{K}_e^0, \qquad p > 1 \,.
\end{equation}
Penalizing the pseudo-density with the power $p$ steers the optimization towards a discrete 0-1 design. The minimum density $\rho_{\mathrm{min}} > 0$ ensures a non-singular global stiffness matrix.

To solve the minimization problem, the Lagrangian $\mathcal{L}$ for the discretized body is formulated by appending the equilibrium and volume constraints via the Lagrange multipliers $\hat{\mathbf{u}}_e$ and $\Lambda$, respectively:
\begin{equation}
    \begin{aligned}
    \label{Lagrangian}
        \mathcal{L} &= \sum_e \mathbf{u}_e^{\mathrm{T}} \mathbf{f}_e - \sum_e \hat{\mathbf{u}}_e^{\mathrm{T}} \left( \mathbf{K}_e(\rho_e) \mathbf{u}_e - \mathbf{f}_e \right) \\
        &\quad + \Lambda \left( \sum_e \rho_{e} V_e - V_{\mathrm{mat}} \right) \,.
    \end{aligned}
\end{equation}

We can identify $\hat{\mathbf{u}}_e = \mathbf{u}_e$ by taking derivative of \eqref{Lagrangian} with respect to $\mathbf{u}_e$.
To find the optimum, we take the derivative with respect to the density $\rho_e$ and set the result to zero :
\begin{equation}\label{drho}
	\frac{\partial \mathcal{L}}{\partial \rho_e} = -\mathbf{u}_e^{\mathrm{T}} \frac{\partial \mathbf{K}_e(\rho_e)}{\partial \rho_e} \mathbf{u}_e + \Lambda V_e = 0 \,.
\end{equation}
With $\partial \mathbf{K}_e / \partial \rho_e = (p/\rho_e)\mathbf{K}_e$, this leads to the optimality condition $B_e = 1$, where $B_e$ is defined as:
\begin{equation}\label{B_dach}
	B_e \coloneqq \frac{\Lambda^{-1}}{V_e} \frac{p}{\rho_{e}} \mathbf{u}_e^{\mathrm{T}} \mathbf{K}_e \mathbf{u}_e \,.
\end{equation}

\subsection{The optimality criteria method (OCM)}

Since this condition cannot be solved analytically for the entire domain, an iterative heuristic update scheme is applied for each iteration step $k$.
With the lower bound $L_e = \max\{(1 - \zeta)\rho_{e,k}, \rho_\mathrm{min}\}$ and the upper bound $U_e = \min\{(1 + \zeta) \rho_{e,k}, 1\}$, the update scheme can be written compactly as:
\begin{equation}\label{update_continuum}
    \rho_{e, k+1} =
    \begin{cases}
        L_e & \text{if } \quad \rho_{e,k} (B_{e,k})^\eta \leq L_e\,, \\
        U_e & \text{if } \quad \rho_{e,k} (B_{e,k})^\eta \geq U_e\,, \\
        \rho_{e,k} (B_{e,k})^\eta & \text{otherwise} \,.
    \end{cases}
\end{equation}

The damping parameter $\eta$ and the move limit $\zeta$ restrict the change in density per step to ensure algorithmic stability \citep{bendsoe2004topology}. Finally, the Lagrange multiplier $\Lambda$ is determined in each iteration, e.g., by means of a bisection scheme, such that the discrete volume constraint $\sum_e \rho_{e, k+1} V_e = V_{\mathrm{mat}}$ is satisfied.

\subsection{Alternative formulation of the optimization problem}

Clapeyron's theorem applies to linearly elastic structures and states:
\begin{equation}
	a(\mathbf{u},\mathbf{u}) = l(\mathbf{u}) \,.
\end{equation}
Thus, the optimization problem \eqref{eq:minimierungsproblem} can be rewritten in terms of the potential energy as follows:
\begin{equation}\label{eq:alternativ_1}
	\begin{aligned}
		\sup_{\mathbf{u}, \mathbf{K} \in \mathcal{K}_\mathrm{adm}} \quad & \frac{1}{2} a(\mathbf{u},\mathbf{u}) -l(\mathbf{u}) \\
		\text{s.t.} \quad & a(\mathbf{u},\mathbf{v}) = l(\mathbf{v}) \quad \forall \mathbf{v} \,.
	\end{aligned}
\end{equation}
This can be equivalently formulated as:
\begin{equation}\label{eq:alternativ}
	\begin{aligned}
		\sup_{\mathbf{K} \in \mathcal{K}_\mathrm{adm}} \left[ \inf_{\mathbf{u}} \frac{1}{2} a(\mathbf{u},\mathbf{u}) -l(\mathbf{u}) \right] \,.
	\end{aligned}
\end{equation}
This formulation, in combination with the SIMP method (cf. Eq. \eqref{SIMP}), allows the optimization problem to be interpreted as a mixed problem (cf. Section \ref{LBB}).

\section{Existing explanations of checkerboarding}\label{literature_chapter}
In this section, we critically review the two main directions for explaining checkerboarding that can be found in the literature.

\subsection{Mixed-field incompatibility}\label{LBB}

\citet{jog1996stability} interpret the topology optimization problem as a mixed variational problem. To understand the idea, we briefly return to the continuous setting. When combining the continuous analogue of \eqref{eq:alternativ} and the continuous SIMP penalization \eqref{SIMP}, the optimization problem takes the following form:
\begin{equation}\label{eq:alternativ2}
	\begin{aligned}
		\sup_{\rho \in [\rho_{\mathrm{min}}, 1]} \left[ \inf_{u \in U} \frac{1}{2} a(u,u) - l(u) \right] \\
		\text{s.t.} \quad \int_\Omega \rho(x) \, \mathrm{d}\Omega = V_{\mathrm{mat}} \,,
	\end{aligned}
\end{equation}
where $a(u,u)$ and $l(u)$ now denote the continuous bilinear and linear forms, respectively, and $U$ is the space of kinematically admissible displacements. Furthermore, the following function space is defined:
\begin{equation}\label{V_p}
	V_\rho = \left\{q \in L^2(\Omega): \int_\Omega q \, \mathrm{d}\Omega = 0 \right\} \,.
\end{equation}
Equipped with the $L^2$-norm and the $L^2$-inner product, this forms a Hilbert space.

The optimality conditions are obtained through variations with respect to the density $\rho$ and the displacement $u$. The resulting equations can be linearized and solved iteratively \citep{jog1996stability}. The resulting discretized system of equations for the iterative update step is given by:
\begin{equation}\label{inc}
	\begin{pmatrix}
		\mathbf{A} & \mathbf{B}^\mathrm{T} \\
		\mathbf{B} & \mathbf{H}
	\end{pmatrix}
	\begin{pmatrix}
		\Delta \mathbf{u} \\ \Delta \boldsymbol{\rho}
	\end{pmatrix}
	=
	\begin{pmatrix}
		\mathbf{f} \\
		\mathbf{g}
	\end{pmatrix} \,.
\end{equation}
Here, $\Delta \mathbf{u}$ and $\Delta \boldsymbol{\rho}$ are the iterative increments. The matrix $\mathbf{A}$ corresponds to the standard structural stiffness matrix, $\mathbf{H}$ represents the Hessian of the objective function with respect to the element densities, and $\mathbf{B}$ is the coupling matrix containing the mixed second derivatives. The right-hand side vectors $\mathbf{f}$ and $\mathbf{g}$ denote the current residuals of the equilibrium and optimality conditions, respectively.

\subsubsection{SIMP with \texorpdfstring{$p=1$}{p=1}}
In \eqref{inc}, the Hessian matrix $\mathbf{H}$ represents the second derivatives with respect to the element densities. Since the stiffness scales with $\rho^p$, these second derivatives vanish ($\mathbf{H}=\mathbf{0}$) if $p=1$, and \eqref{inc} corresponds to the form of the discrete incompressible Stokes problem. For any nonlinear penalization ($p \neq 1$), the second derivatives are non-zero ($\mathbf{H} \neq \mathbf{0}$) \citep{jog1996stability}.

To satisfy the global volume constraint, it is necessary that $\Delta \rho \in V_\rho$. However, this condition on the increment $\Delta\rho$ is not sufficient due to the physical box constraints of the variables. The updated density must remain within the admissible bounds $[\rho_{\mathrm{min}}, 1]$, which requires:
\begin{equation}
	\rho_{\mathrm{min}} - \rho_k \leq \Delta\rho \leq 1 - \rho_k \,,
\end{equation}
where $\rho_k$ denotes the density at the beginning of the current increment. Because of these strict upper and lower bounds, the set of admissible increments $\Delta\rho$ is not closed under scalar multiplication. Scaling a valid, small increment by an arbitrarily large factor will inevitably violate the physical bounds. Consequently, this restricted set does not constitute a vector space, and therefore cannot form a Hilbert space.

In a discrete mixed formulation such as \eqref{inc}, a finite element pair denotes the combination of interpolation spaces chosen for the incremental displacement and density fields. Unstable element pairs violate the LBB conditions \citep{brezzi1991mixed} and can produce spurious oscillations.

Since the LBB conditions explicitly require a Hilbert space, they cannot be applied for the analysis of \eqref{inc}. Therefore, \citet{jog1996stability} rely on numerical eigenvalue analyses to detect zero eigenvalues. By demonstrating that the system matrix is not full-rank, they numerically prove the existence of an instability. This instability also occurs in the classical Stokes problem, where unstable discretizations that do not satisfy the LBB conditions trigger spurious pressure modes. The corresponding checkerboard patterns yield zero internal work and thus remain "unseen" by the displacement field, polluting the entire solution \citep{sani1981cause}.

\subsubsection{SIMP with \texorpdfstring{$p>1$}{p>1}}
If $p \neq 1$, then $\mathbf{H} \neq \mathbf{0}$. In this case, the system matrix no longer conforms to the structure of the discrete incompressible Stokes problem. Nevertheless, to demonstrate the instability of certain element pairs, \citet{jog1996stability} conduct numerical eigenvalue experiments for this case as well. Their analyses reveal that, similar to the case with $p=1$, the system matrix exhibits zero eigenvalues for specific element pairs. This lack of full rank indicates the presence of spurious modes and hence the existince of a numerical instability.

\subsubsection{Applicability to the OCM}
The update scheme on which the OCM is based evaluates displacement and density sequentially rather than simultaneously. Ideally, one would validate the assumption of mixed-field incompatibility by directly comparing the checkerboard modes obtained from a mixed finite element formulation with those generated by the OCM. However, the mixed formulation analyzes a coupled system matrix that becomes rank-deficient and triggers zero-energy modes. In contrast, the decoupled OCM solves the equilibrium using a standard, full-rank stiffness matrix. Thus, the numerical instability of the mixed formulation does not directly translate to the OCM formulation.

Furthermore, \citet{jog1996stability} state that the system matrix in the mixed formulation only approaches the singular state, thus triggering the instability, as the optimization advances and elements reach the discrete density bounds of $\rho_\mathrm{min}$ and $1$. Numerical observations with the OCM, however, contradict this chronological sequence. As illustrated in the next section in the intermediate iterations of the optimization process (see for example Fig. \ref{beam_p3_15}), checkerboard patterns already manifest clearly while intermediate densities still dominate the solution over the domain. This raises the doubts on whether the instability observed in the mixed formulation is really the underlying mathematical mechanism of checkerboarding, or rather a consequence of the pattern itself.

Another critical discrepancy is the spatial localization of the patterns. If checkerboarding in topology optimization was a purely mathematical spatial instability, analogous to spurious pressure modes in the discrete Stokes problem, one would expect the alternating pattern to propagate continuously throughout the domain. In the Stokes problem, an unstable element pair typically corrupts the entire pressure field \citep{boffi2013mixed, sani1981cause}. In contrast, OCM checkerboard patterns are highly localized. They do not uniformly pollute the domain, but strictly cluster in specific regions determined by the global boundary conditions, leaving other areas completely free of numerical artifacts. However, the analysis framework based on the LBB conditions provides no rationale for this localized spatial distribution.

As a final remark, an interesting nuance arises when examining the stability of specific element pairs reported by \citet{jog1996stability}. In the classical Stokes problem, utilizing biquadratic velocity interpolation with piecewise constant pressure ($Q_9/P_0$) is known to be susceptible to spurious pressure modes \citep{boffi2013mixed, donea2003finite}. Yet, \citet{jog1996stability} observe that this element pair remains stable in their topology optimization framework ($Q_9$ displacement with uniform density). This observation suggests that the instability might not stem from a generalized mixed-field incompatibility. Instead, it indicates that the checkerboard issue is closely tied to the kinematic limitations of bilinear elements themselves, making a purely mechanical interpretation a plausible complementary perspective.

\subsection{Locking}\label{locking}

\citet{diaz1995checkerboard} approach the checkerboard issue from a mechanical perspective, demonstrating that bilinearly interpolated finite elements artificially overestimate the stiffness of 2D checkerboard layouts.

\subsubsection{Definitions}

In topology optimization, regions with strong and weak material exist within the optimized structure. Let $E^0$ be the elasticity tensor of the solid material, while the stiffness of the weak material is assumed to be zero.

If the material is arranged in a 2D checkerboard pattern, it can be interpreted as a periodic microstructure for which an effective macroscopic stiffness tensor can be derived \citep{diaz1995checkerboard}. Let $\overline{E}_{4}$ denote the effective stiffness of a checkerboard pattern modeled with bilinear $Q_4$ elements, and $\overline{E}_9$ the effective stiffness when modeled with biquadratic $Q_9$ elements. For a given macroscopic strain field $\overline{\epsilon}$, the internal strain energy densities are given by:
\begin{equation}
	w_4 = \frac{1}{2}\overline{E}_{4} \overline{\epsilon} \cdot \overline{\epsilon}\,, \qquad 
	w_9 = \frac{1}{2}\overline{E}_{9} \overline{\epsilon} \cdot \overline{\epsilon} \,.
\end{equation}

To evaluate these energies against physically optimal microstructures, they are compared to the internal energy of an optimal rank-2 layered material aligned with the principal strains at a volume fraction of $\rho = 1/2$:
\begin{equation}
	w_\mathrm{Rank2} = \max_{\text{Rank-2}, \,\rho= 1/2}\frac{1}{2} \overline{E}_\mathrm{Rank2} \overline{\epsilon} \cdot \overline{\epsilon} \,.
\end{equation}

Furthermore, to account for the SIMP penalization with penalty factor $p$, the internal energy of a material with a uniformly distributed intermediate density of $\rho = 1/2$ is defined as:
\begin{equation}
	w_\mathrm{Const} = \frac{1}{2}\left(\frac{1}{2}\right)^p E^0\overline{\epsilon}\cdot\overline{\epsilon} \,.
\end{equation}

\subsubsection{SIMP with \texorpdfstring{$p=1$}{p=1}}\label{diazp1}
\citet{diaz1995checkerboard} mathematically proved that without SIMP penalization ($p=1$), the following inequalities hold for all prescribed multiaxial strain fields $\overline{\epsilon}$ in a 2D problem:
\begin{equation}\label{w4wRank2}
	w_4 \geq w_\mathrm{Rank2} \quad \text{and} \quad w_9 < w_\mathrm{Rank2} \,.
\end{equation}

\begin{figure}[htbp]
	\centering
	\includegraphics[width=0.95\columnwidth]{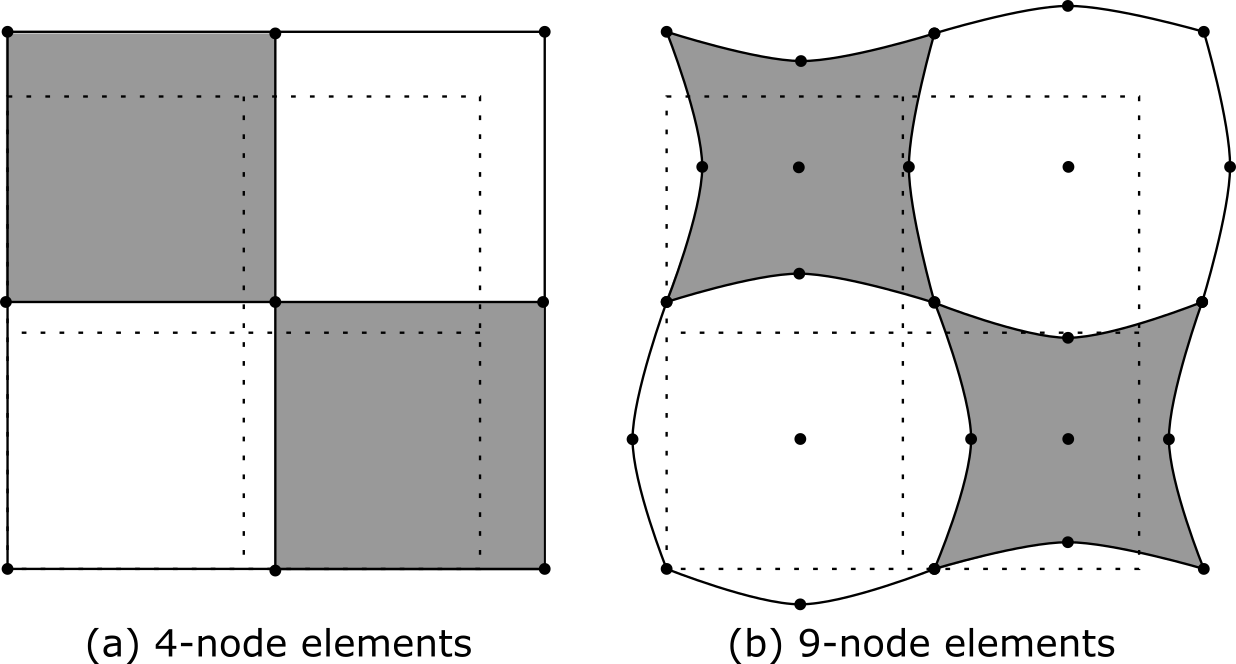}
	\caption{Deformation of a periodic checkerboard microstructure under macroscopic strain, illustrating the artificial stiffness of $Q_4$ elements (a) versus the physical pinching enabled by $Q_9$ elements (b) \citep{diaz1995checkerboard}.}
	\label{Lock}
\end{figure}

Figure \ref{Lock} illustrates the underlying mechanism of the artificial stiffness in bilinear $Q_4$ elements. It is instructive to compare this phenomenon to classical shear locking, a well-documented numerical artifact in standard displacement-based finite element formulations \citep{bathe2014finite, hughes1987finite}. Both artifacts originate from the same fundamental kinematic restriction of the bilinear $Q_4$ element: its inability to model curved edges. However, while classical shear locking stiffens an individual element under a macroscopic bending load due to the introduction of artificial shear energy for actual bending modes, checkerboard locking occurs due to the inability to sense the weakness of the neighboring elements. With biquadratic $Q_9$ elements, the structural layout allows for local necking (pinching) at the shared element edges, reflecting the physical weakness of a single-node connection. Bilinear elements, however, kinematically do not allow for this necking, which allows them to transport artificially high loads through the hinged connections on the nodes.

Equation \eqref{w4wRank2} demonstrates that a checkerboard pattern modeled with bilinear elements appears stiffer than the analytically optimal rank-2 material at the same density. While this proves that the $Q_4$ pattern exhibits an artificially stiff behavior, it alone does not justify why the optimization algorithm actively creates the checkerboard. The actual question would be whether that pattern is also stiffer than other $Q_4$-discretized configurations, notably a continuous homogeneous material distribution. 
For $p=1$, one can show that \citep{diaz1995checkerboard}:
\begin{equation}\label{w4wconst}
    w_4 = w_\mathrm{Const} \,.
\end{equation}
The checkerboard thus has the same stiffness as the uniformly distributed material with half of the density, which implies that the choice between checkerboard and gray is arbitrary from an asymptotic point of view.

A further point to consider is the limit case in which the first relation in \eqref{w4wRank2} becomes an equality. In a purely uniaxial stress state, where the ratio of principal strains strictly corresponds to the Poisson ratio ($\epsilon_{\mathrm{II}}/\epsilon_{\mathrm{I}} = -\nu$), the optimal rank-2 microstructure mathematically degenerates into a rank-1 material and we thus obtain:
\begin{equation}\label{eq4R2}
    w_4 = w_\mathrm{Rank2} \,.
\end{equation}

Note that this implies that, from a microscopic perspective, the checkerboard pattern under purely uniaxial load is exactly as stiff as continuous struts aligned in the direction of the non-zero principal stress. 

\subsubsection{SIMP with \texorpdfstring{$p>1$}{p>1}}
When the SIMP penalization is active ($p>1$), the following relations apply \citep{diaz1995checkerboard}:
\begin{equation}\label{ungl_4}
	w_4 > w_\mathrm{Const} \,,
\end{equation}
\begin{equation}\label{ungl_9}
	w_9 < w_\mathrm{Const} \quad \text{for} \quad p < p^*_1 \,. 
\end{equation}

For bilinear elements, \eqref{ungl_4} is the driving force behind checkerboard formation. The SIMP method penalizes intermediate uniform densities, so the optimizer seeks discrete 0-1 configurations. Since the bilinear checkerboard has the same stiffness as a gray continuous material (see \eqref{w4wconst}), it is the algorithmic response to gray domains in the non-penalized setup.

Conversely, \eqref{ungl_9} shows that biquadratic elements do not suffer from this artificial stiffness. While the critical penalty value $p^*_1$ is typically smaller than the standard practical values used in topology optimization (e.g., $p=3$), this relation establishes a fundamental baseline: even with some penalization, the checkerboard is already weaker than the uniform gray distribution. Consequently, since continuous solid regions are entirely unaffected by the penalty and out-stiff intermediate gray densities, they are superior to a checkerboard arrangement consisting of the same material. Therefore, biquadratic discretizations correctly identify the structural weakness of the checkerboards and naturally suppress the creation of such patterns.

\subsubsection{Missing link to spatial variability and localization}\label{Zusammenfasssung_Schach}
The microstructural analysis by \citet{diaz1995checkerboard} provides a compelling mechanics-driven explanation for the checkerboard phenomenon: bilinear elements exhibit an artificially high stiffness due to their inability to sense the physical weakness of connected neighbor elements. As a result, the checkerboard mimics the stiffness of an equivalent domain with non-penalized gray material, a loophole that the OCM algorithm actively exploits to bypass the SIMP penalty.

However, this analysis does not explain the macroscopic spatial variability of checkerboard patterns. It provides a reason why checkerboards are mathematically more attractive to the optimizer than intermediate densities. But it fails to explain why the optimizer does not deploy this artificially stiff layout everywhere across the design domain. In the following, we will focus on investigating this spatial variability. 

\section{Stress-state dependence of checkerboard formation and localization}\label{main}

In the following, we investigate the hypothesis that checkerboard formation is governed by the underlying stress state. Using a series of systematic numerical experiments, we demonstrate that checkerboards localize in regions of multiaxial load transfer, while uniaxial load paths remain checkerboard-free. Based on the mechanical relations established in Section~\ref{literature_chapter}, these observations form the basis for a unified mechanical explanation of checkerboard localization.

\begin{figure*}[htbp]
	\centering
	\includegraphics[width=0.9\textwidth]{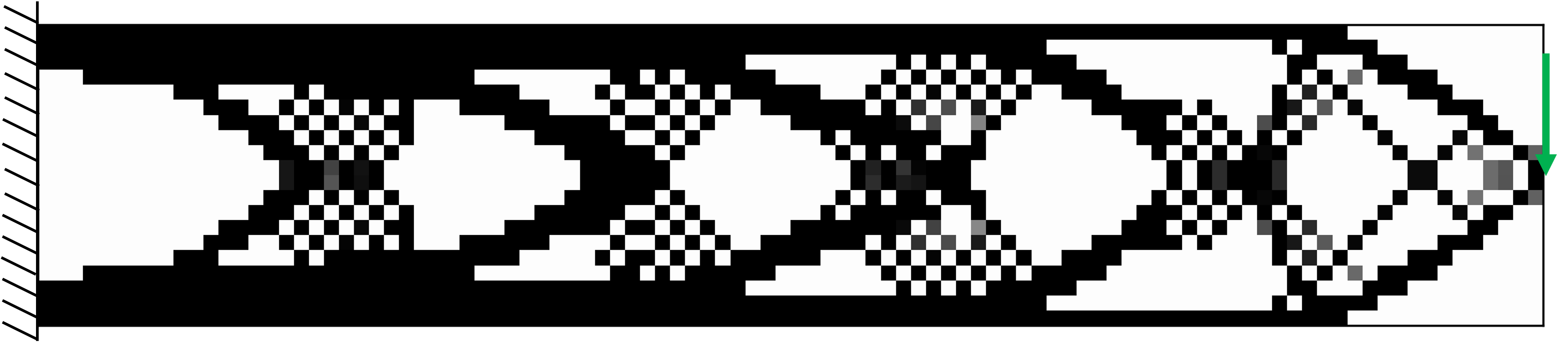}
	\caption{Optimization of a cantilever beam discretized with $Q_4$ elements (Domain: 100x20 elements, volume fraction $v_\mathrm{mat} = 0.5$, penalty $p = 3$).}
	\label{beam_intro_beam}
\end{figure*}

\subsection{Hypothesis and phenomenological observations}

Inspection of classical topology optimization benchmarks immediately reveals that checkerboard patterns do not emerge uniformly throughout the design domain, but localize within specific regions. Figure~\ref{beam_intro_beam} shows a representative cantilever beam optimized using $Q_4$ elements in the absence of filtering and other spatial regularization techniques. The localized nature of the checkerboard pattern is clearly visible.

To interpret this material distribution, it is useful to recall the classical strut-and-tie model \citep{bruggi2009generating}. The bending moment is transferred through continuous tension and compression chords along the outer boundaries, whereas the shear force is carried through a diagonal web system. Remarkably, checkerboard patterns appear predominantly within these shear-transfer regions, while the uniaxial tension and compression chords remain checkerboard-free.

This observation suggests that checkerboard formation is governed by the underlying stress state. In particular, we hypothesize that checkerboards emerge in regions requiring multiaxial load transfer, whereas predominantly uniaxial load paths naturally favor continuous solid material distributions.

While checkerboards visually coincide with regions of pronounced macroscopic shear, the corresponding shear stress components depend on the chosen coordinate system and therefore do not provide an objective measure for predicting their occurrence. To characterize the local stress state, we therefore introduce the invariant multiaxiality index $\xi$, defined for two-dimensional plane-stress problems as
\begin{equation}\label{eq:xi}
	\xi \coloneqq 
	\begin{cases}
		\sigma_{\mathrm{II}}/\sigma_{\mathrm{I}} & \text{if } |\sigma_{\mathrm{I}}| > 10^{-12}\,, \\
		0 & \text{otherwise} \,.
	\end{cases}
\end{equation}
The principal stresses are ordered such that $|\sigma_I|\geq|\sigma_{\textit{II}}|$, implying $\xi\in[-1,1]$. Values of $\xi$ close to zero indicate predominantly uniaxial stress states, whereas increasing magnitudes of $\xi$ correspond to progressively stronger multiaxial loading.

To establish causality rather than mere correlation, it is insufficient to demonstrate that multiaxial stress states are present in regions where checkerboards have already formed. Instead, the stress state must be examined during the early stages of the optimization process, prior to the onset of the instability. The following benchmark studies are designed to test this hypothesis and to identify the conditions under which checkerboard patterns emerge.

\subsection{Numerical setup}

The numerical investigations are conducted using a density-based topology optimization framework based on the classical 99-line educational code by \citet{sigmund200199}. To enable a comparative analysis of the influence of the displacement interpolation, the framework was extended to support both bilinear ($Q_4$) and biquadratic ($Q_9$) finite elements. To allow checkerboard patterns to develop naturally, all spatial regularization techniques are explicitly deactivated. The prescribed volume fractions and SIMP penalty factors are specified individually for each benchmark problem.

Unless stated otherwise, all computations employ a Young's modulus of $E=33,000$, a Poisson ratio of $\nu=0.2$, OCM parameters $\zeta=0.2$ and $\eta=0.5$, and a minimum density of $\rho_\mathrm{min}=0.01$. The optimization process is terminated when the maximum density change between two consecutive iterations falls below $10^{-3}$ or after a maximum of 99 iterations. In many cases, the iteration limit is reached because checkerboard patterns induce persistent iteration-to-iteration oscillations of the density field.


\subsection{Uniaxial load transfer: absence of checkerboarding}\label{sec:tension_rod}

We first investigate whether checkerboard patterns can emerge in the absence of multiaxial stress states. To this end, a tension rod is considered whose boundary conditions are chosen such that the resulting stress field remains predominantly uniaxial throughout the domain. The left boundary is constrained by roller supports, allowing free lateral contraction, while a single node is fully fixed to prevent rigid body motion. The load is applied through a kinematically constrained subset of the right boundary, forcing these nodes to displace equally in the horizontal direction.

\begin{figure}[t!]
\centering
\includegraphics[width=0.95\columnwidth]{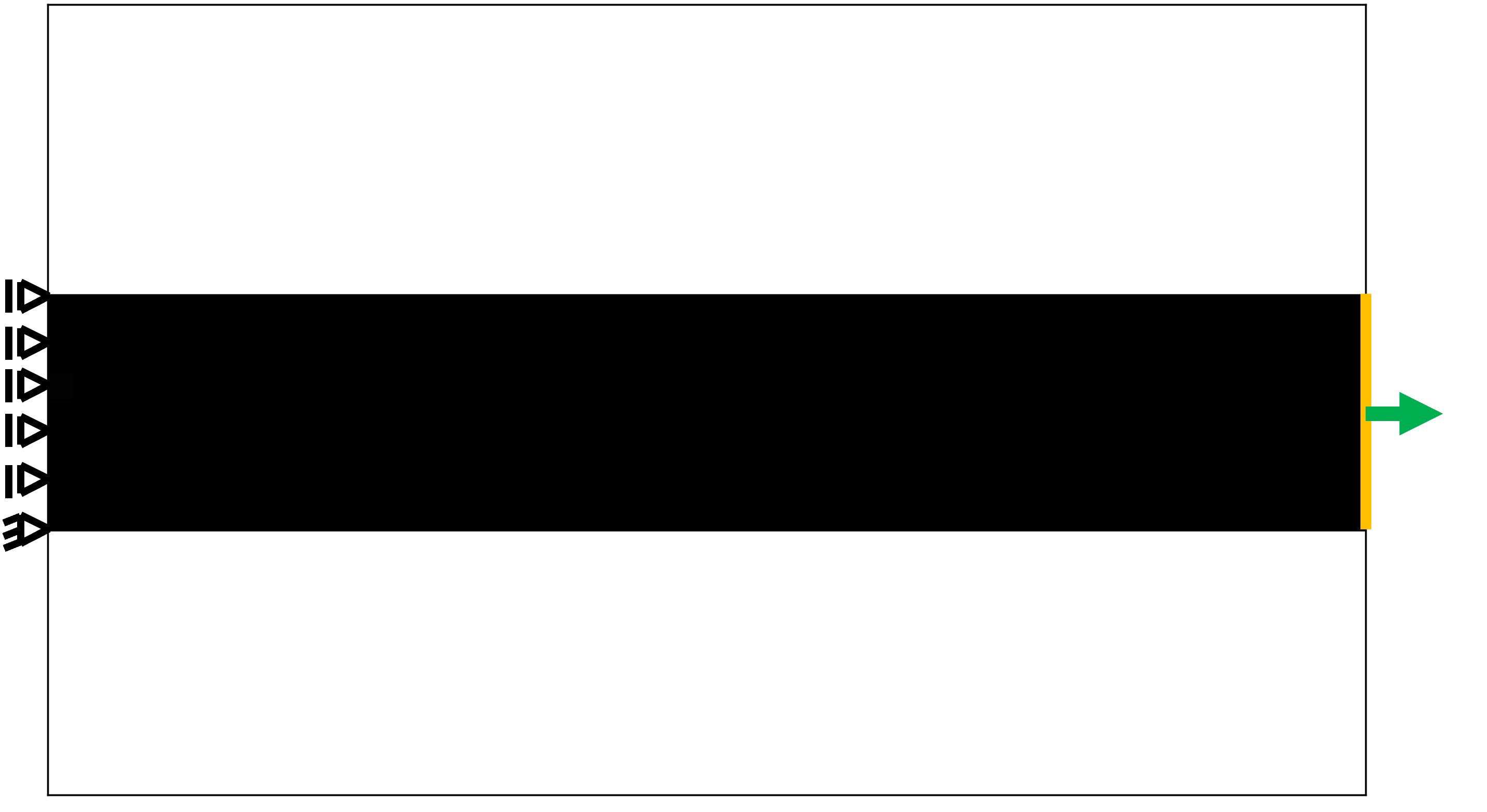}
\caption{Uniaxially loaded bar; boundary constraint matching the volume capacity (50$\times$30 $Q_4$ elements, volume fraction $v_\mathrm{mat}=0.3$, SIMP penalty factor $p=3$).}
\label{uniax_30}
\end{figure}

In the first experiment, shown in Figure~\ref{uniax_30}, the width of the constrained boundary segment is chosen to exactly match the final continuous solid strut result, to which the optimizer converges. No checkerboard patterns are observed. 

Since a checkerboard consists of approximately 50\,\% void space, it requires a substantially larger geometric cross-section to accommodate the same amount of material. Forming such a porous layout would widen the load transfer and reduce structural stiffness. A checkerboard therefore cannot compete with a continuous strut when the available geometric width is fully utilized by the material volume. 

This observation can be further corroborated via relations~\eqref{w4wconst} and~\eqref{eq4R2}. Together, they imply that under purely uniaxial loading a checkerboard pattern possesses the same asymptotic stiffness as a continuous solid strut containing the same amount of material. Consequently, checkerboards do not provide any intrinsic stiffness advantage in uniaxial load transfer.

\begin{figure}[t!]
\centering
\includegraphics[width=0.95\columnwidth]{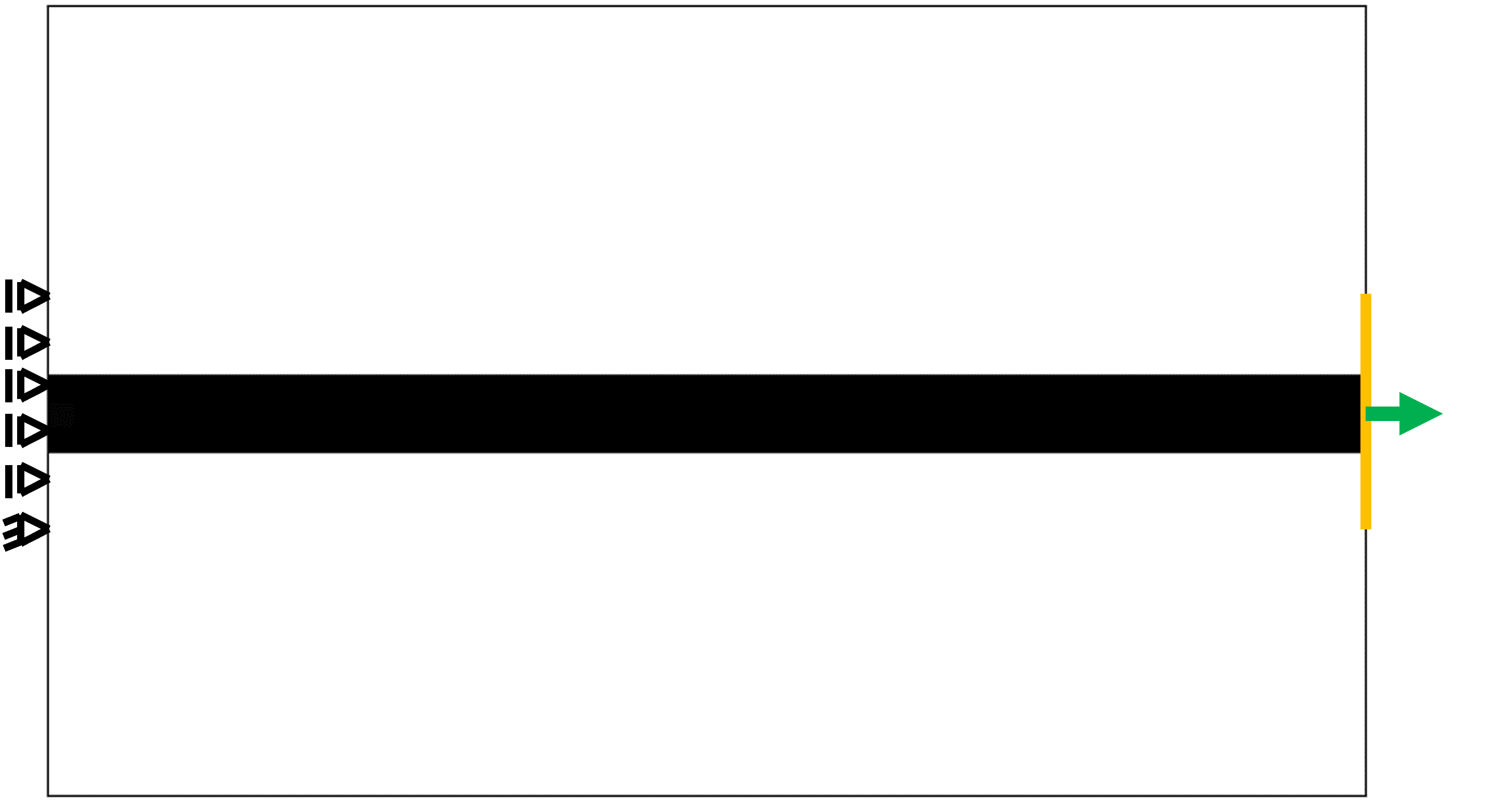}
\caption{Uniaxially loaded bar; boundary constraint narrower than the available volume capacity (50$\times$30 $Q_4$ elements, volume fraction $v_\mathrm{mat}=0.1$, SIMP penalty factor $p=3$).}
\label{uniax_10}
\end{figure}

To verify that the absence of checkerboarding is not merely a consequence of the prescribed boundary width, the benchmark is repeated with a reduced volume fraction. As a result, the width of the constrained boundary segment significantly exceeds that required by the available material volume. The corresponding result in Figure~\ref{uniax_10} shows that despite the additional geometric freedom, the optimizer again converges to a continuous solid strut without checkerboard patterns.

We conclude that the absence of checkerboarding cannot be attributed solely to geometric restrictions. Even when sufficient space is available, the optimizer continues to favor a continuous solid strut. The reason lies in the finite extent of the checkerboard pattern. The stiffness equivalence established by~\eqref{w4wconst} and~\eqref{eq4R2} is an asymptotic result for an infinitely repeating microstructure. In a finite structure, however, the outer boundary of the checkerboard constitutes a pronounced weak region. The boundary nodes cannot transfer load through neighboring void elements and therefore exhibit a substantially reduced load-carrying capacity compared to the interior of the pattern.

Continuous solid struts do not suffer from this boundary weakness. Consequently, while checkerboards and solid struts may become asymptotically equivalent in the limit of infinitely large periodic patterns, finite checkerboards remain mechanically inferior under uniaxial load transfer. 

Taken together, these results indicate that predominantly uniaxial load transfer does not promote checkerboard formation. In such regions, continuous solid load paths remain mechanically preferable, and the optimizer consistently converges to checkerboard-free configurations.

\subsection{Multiaxial load transfer: onset of checkerboarding}\label{sec:tension_rod_hindered}

We now consider a benchmark that deliberately introduces localized multiaxial stress states. To this end, a compression rod is analyzed in which lateral contraction is prevented at the left support. By restraining the lateral deformation associated with the Poisson effect, a localized multiaxial stress region is generated near the left support, whereas the remainder of the structure remains predominantly uniaxially loaded.

\begin{figure*}[t!]
	\centering
	
	\includegraphics[width=0.9\textwidth]{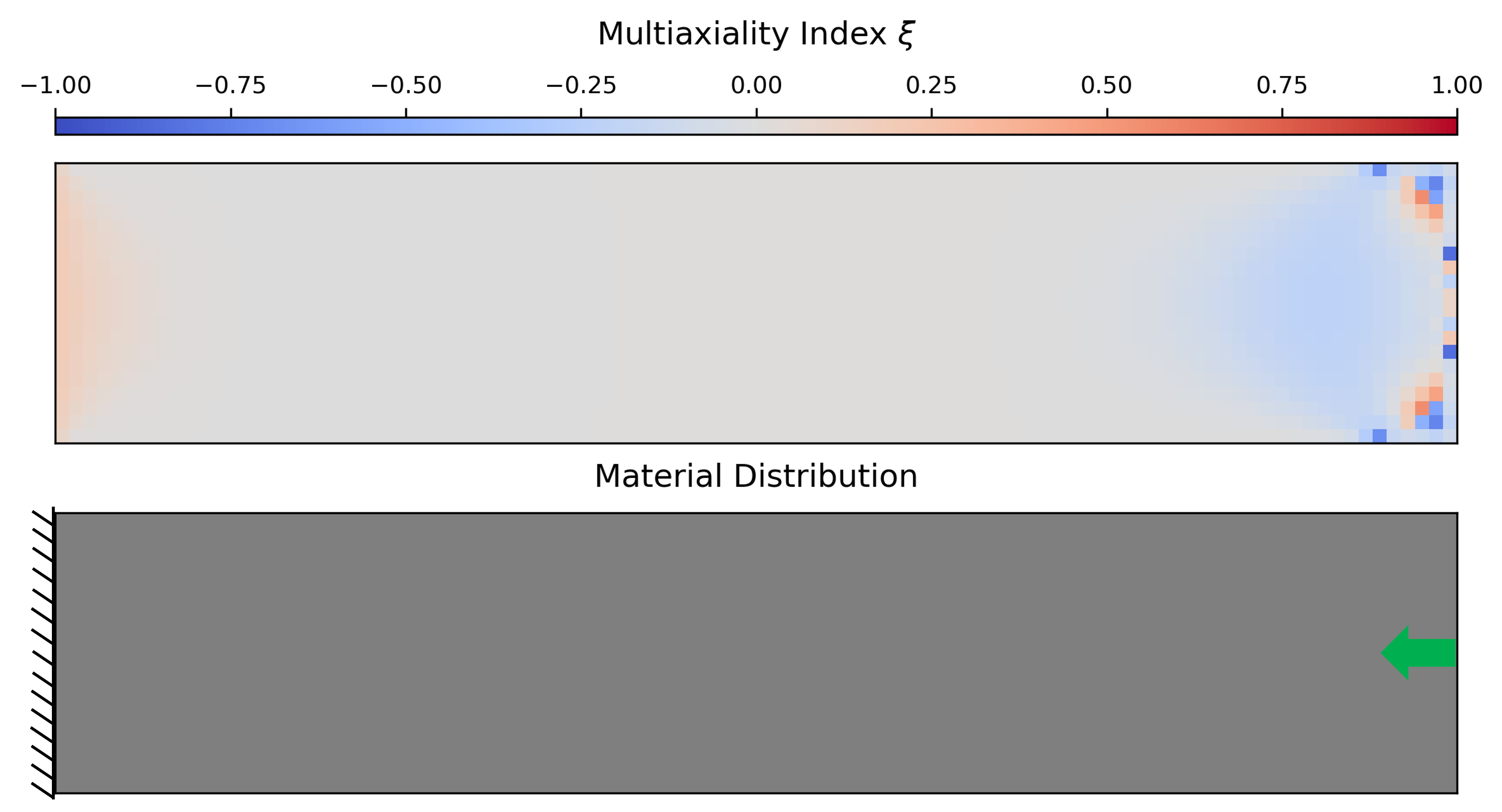}
	\caption{Initial state: compression rod with constrained lateral contraction (100$\times$20 $Q_4$ elements, no SIMP penalization: $p=1$).}
	\label{tension2_1}
	
	\vspace{0.2cm} 
	
	\includegraphics[width=0.9\textwidth]{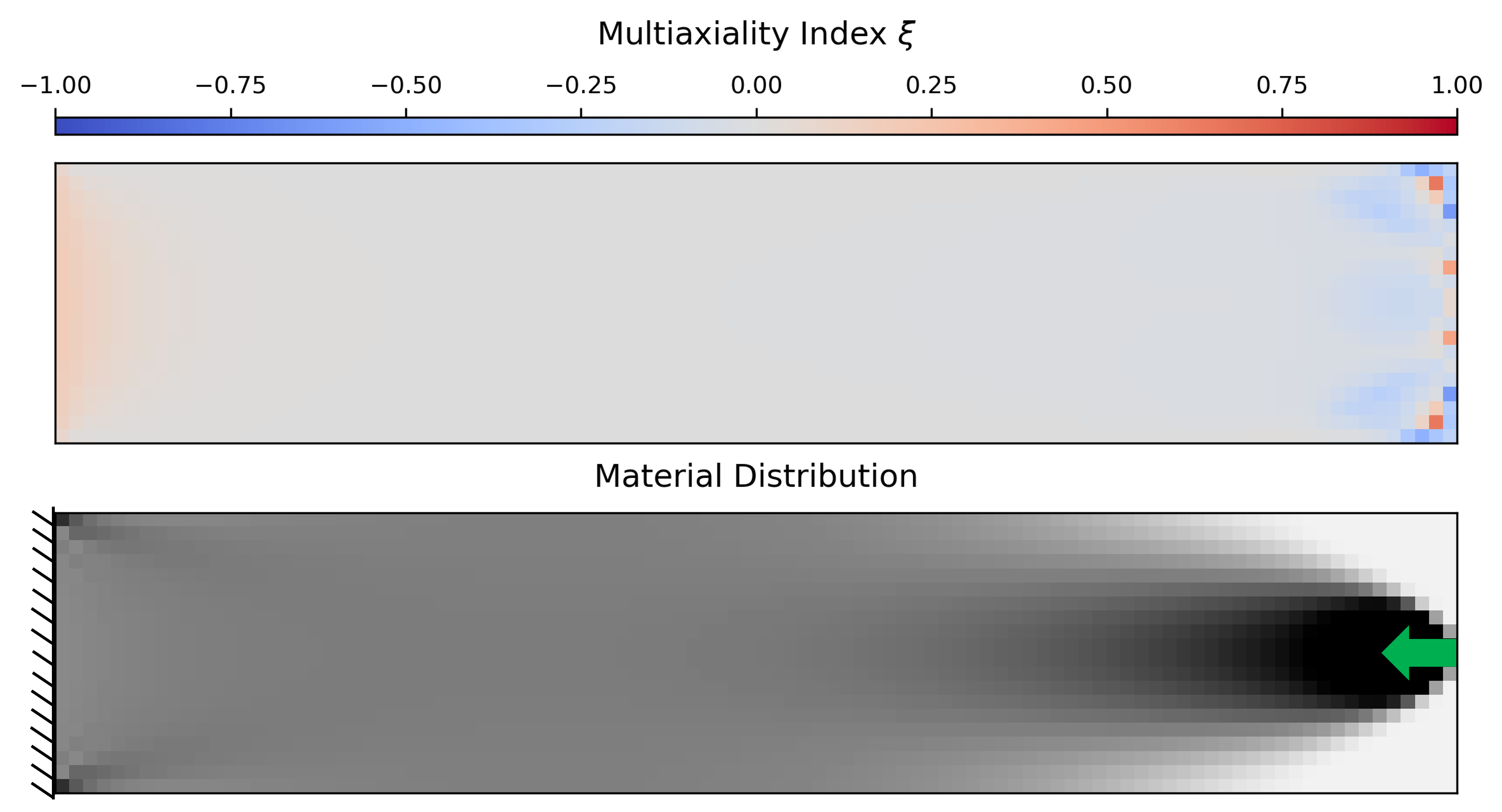}
	\caption{Early optimization stage after iteration 10: onset of local checkerboarding (seeding) in multiaxial stress region at the left boundary (100$\times$20 $Q_4$ elements, no SIMP penalization: $p=1$).}
	\label{tension2_10}
	
\end{figure*}

\begin{figure*}[t!]
	\centering
	\includegraphics[width=0.9\textwidth]{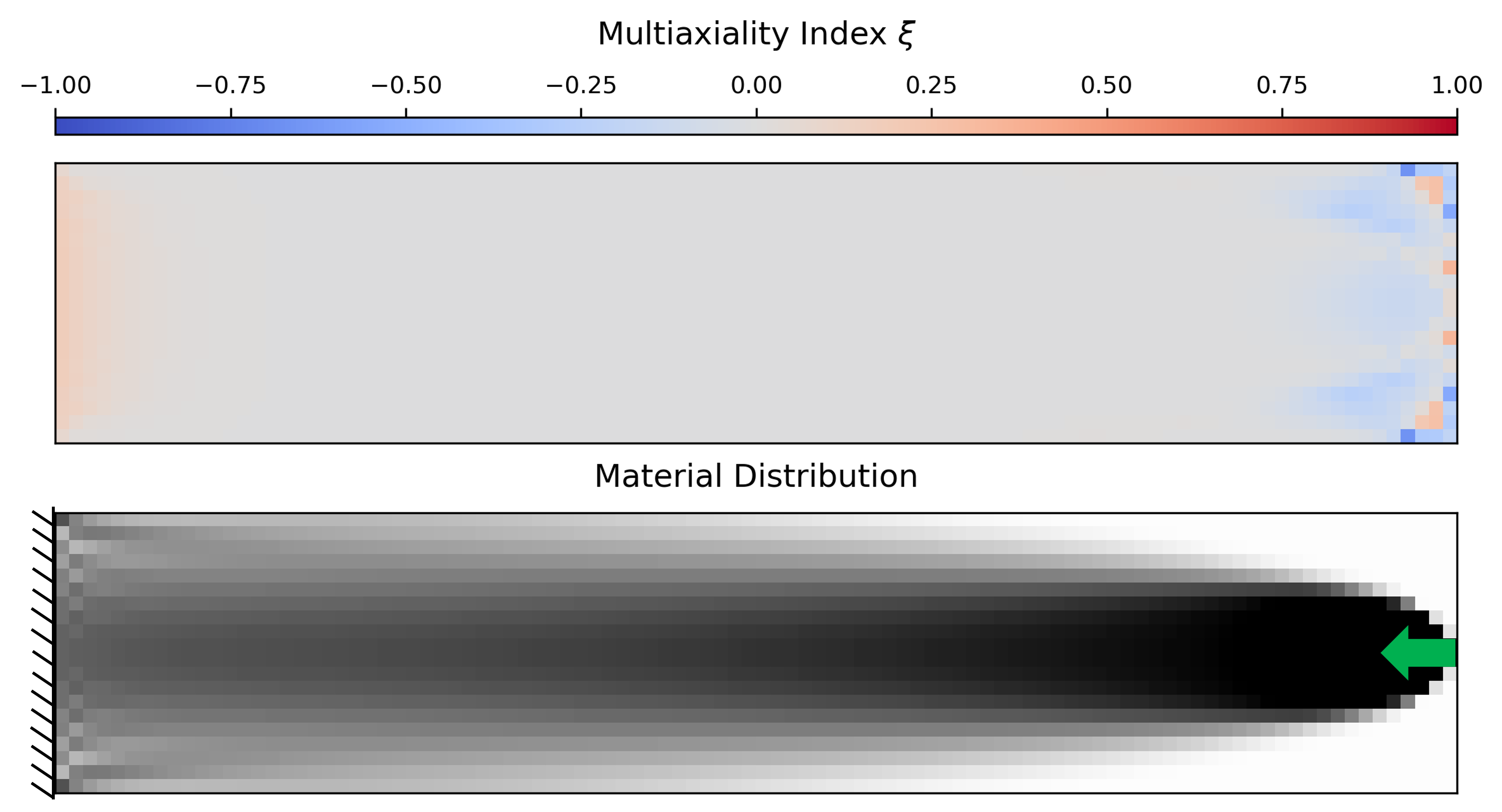}
	\caption{Final topology after iteration 99: checkerboarding at the left support and solid material at the location where the load is applied (100$\times$20 $Q_4$ elements, no SIMP penalization: $p=1$).}
	\label{tension2_99}
\end{figure*}

To expose the underlying load-transfer mechanisms without the influence of SIMP penalization, we first consider the case $p=1$. The evolution of the multiaxiality index $\xi$ and the corresponding density distribution in the initial state, after 10 iterations and at the end of the iterative optimization procedure after 99 iterations are shown in Figures~\ref{tension2_1}, \ref{tension2_10} and \ref{tension2_99}, respectively.

 In Figure~\ref{tension2_1}, we can observe a localized region of elevated multiaxiality adjacent to the laterally constrained support at the left boundary. In the initial state, no checkerboard pattern exists and the density field remains essentially uniform. Consequently, the multiaxial stress state clearly precedes the appearance of potential numerical artifacts.

During the subsequent optimization process, checkerboard patterns begin to emerge precisely within this multiaxial region. Figure~\ref{tension2_10} illustrates an early stage of this process. The pattern is initially triggered by small numerical differences between neighboring element sensitivities, a phenomenon referred to here as \emph{seeding}. Once initiated, however, the checkerboard is actively amplified by the optimization algorithm.

As the optimization progresses, the initially small perturbations evolve into a fully developed checkerboard pattern localized at the constrained left support, see Figure~\ref{tension2_99}. Importantly, the spatial location of the checkerboard coincides with the multiaxial stress region identified at the left boundary by the initial multiaxiality distribution. This observation provides direct evidence that multiaxiality acts as the structural trigger for checkerboard formation.

\begin{figure*}[t!]
	\centering
	
	\includegraphics[width=0.9\textwidth]{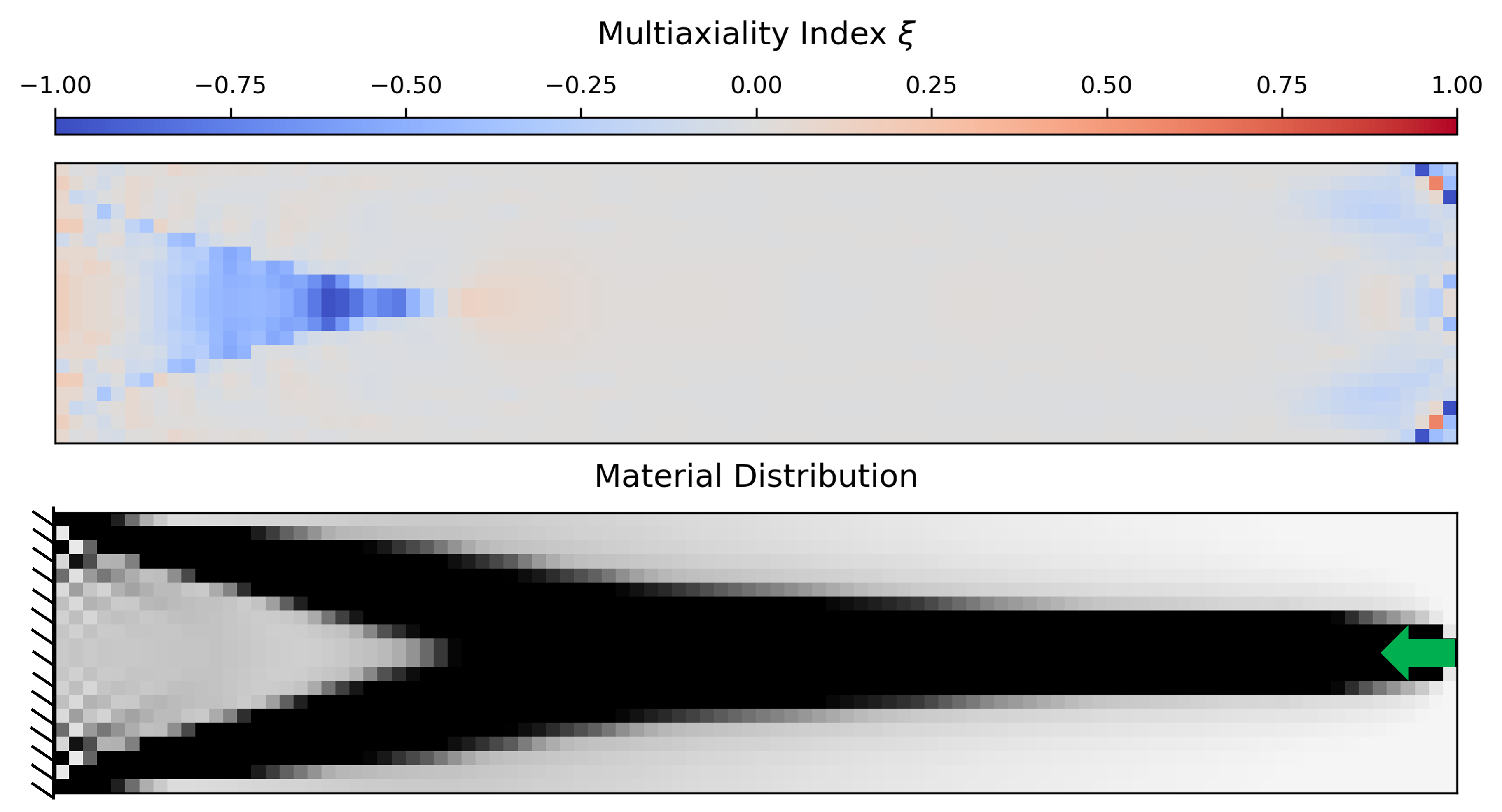}
	\caption{Early optimization stage after iteration 11: onset of checkerboarding at the left boundary (100$\times$20 elements, SIMP penalty factor $p=3$).}
	\label{tension2_11}
	
	\vspace{0.2cm}
	
	\includegraphics[width=0.9\textwidth]{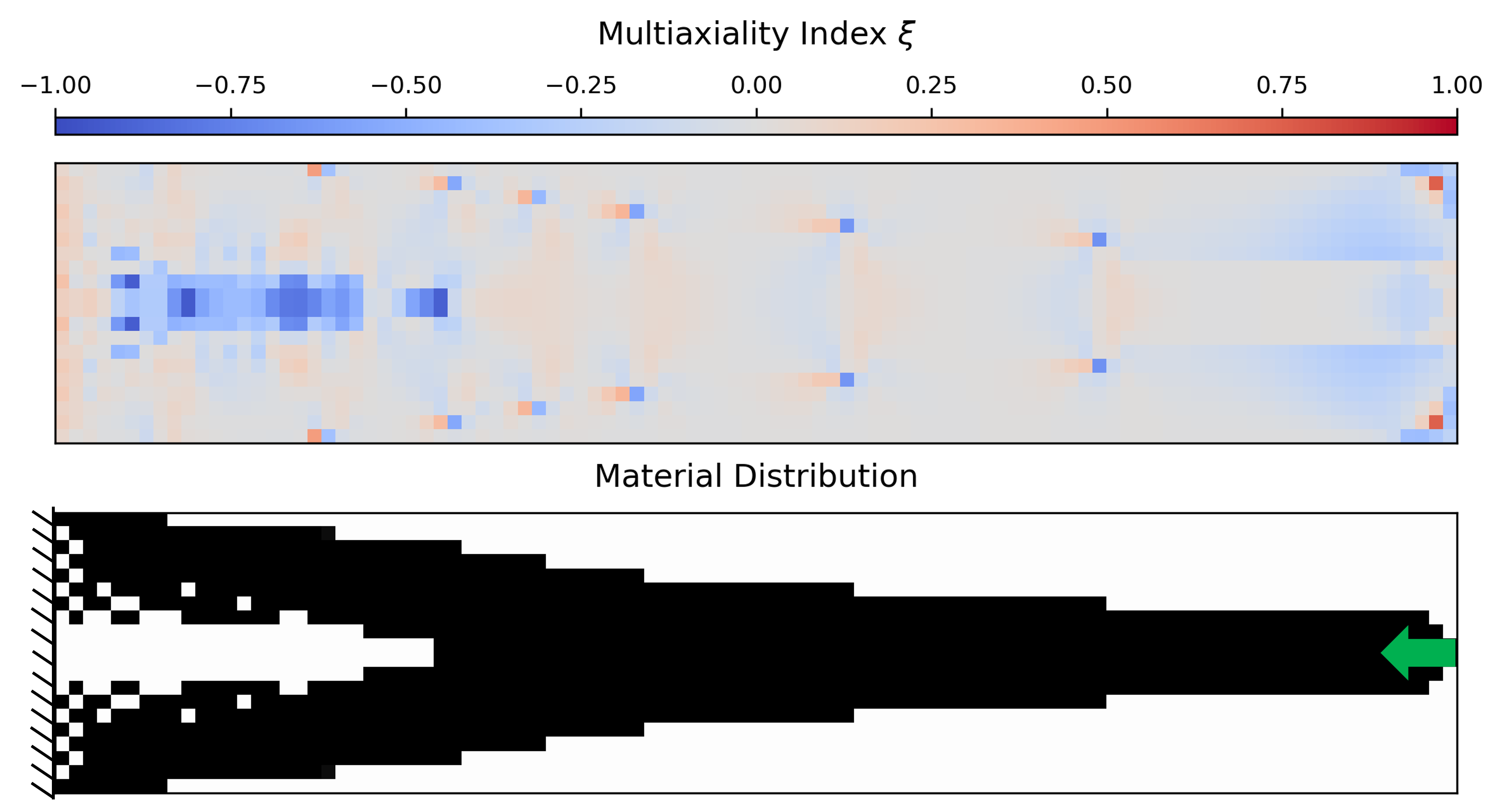}
	\caption{Final topology after iteration 44: checkerboarding at the left support and solid struts at locations of uniaxial stress state (100$\times$20 $Q_4$ elements, SIMP penalty factor $p=3$).}
	\label{tension2_44}
	
\end{figure*}

To investigate the influence of SIMP penalization, the same benchmark is repeated with the penalty factor $p=3$. The corresponding results are shown in Figures~\ref{tension2_11} and~\ref{tension2_44}.

We can observe that the introduction of SIMP penalization does not alter the qualitative behavior. Checkerboard patterns again emerge in the vicinity of the multiaxial support region and remain absent in the region of predominantly uniaxial load transfer. 
Furthermore, the final checkerboard arrangement is not perfectly alternating but contains local imperfections and adjacent void clusters. These features originate during the iterative optimization procedure and remain trapped as local minima. Their existence provides an additional indication that checkerboard formation cannot be interpreted as a purely mathematical instability mode analogous to the globally oscillatory pressure modes encountered in unstable discretizations of the incompressible Stokes problem.

\subsection{Cantilever beam: localization in multiaxial shear regions}\label{sec:cantilever_beam}

\begin{figure*}[t!]
	\centering
	
	\includegraphics[width=0.9\textwidth]{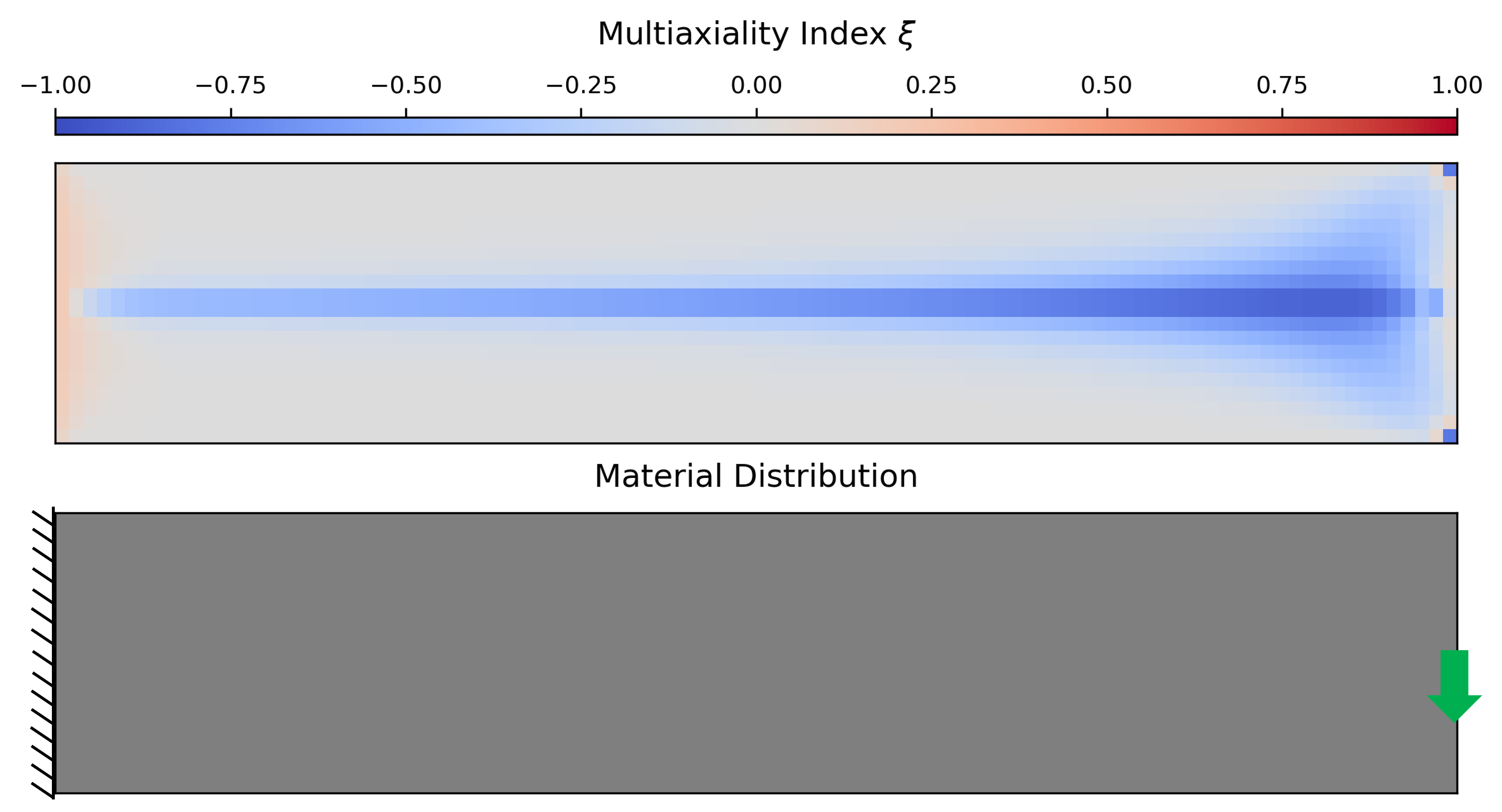}
	\caption{Initial state of the cantilever beam (100$\times$20 $Q_4$  elements, no SIMP penalization: $p=1$).}
	\label{beam_p1_0}
	
	\vspace{0.2cm} 
	
	\includegraphics[width=0.9\textwidth]{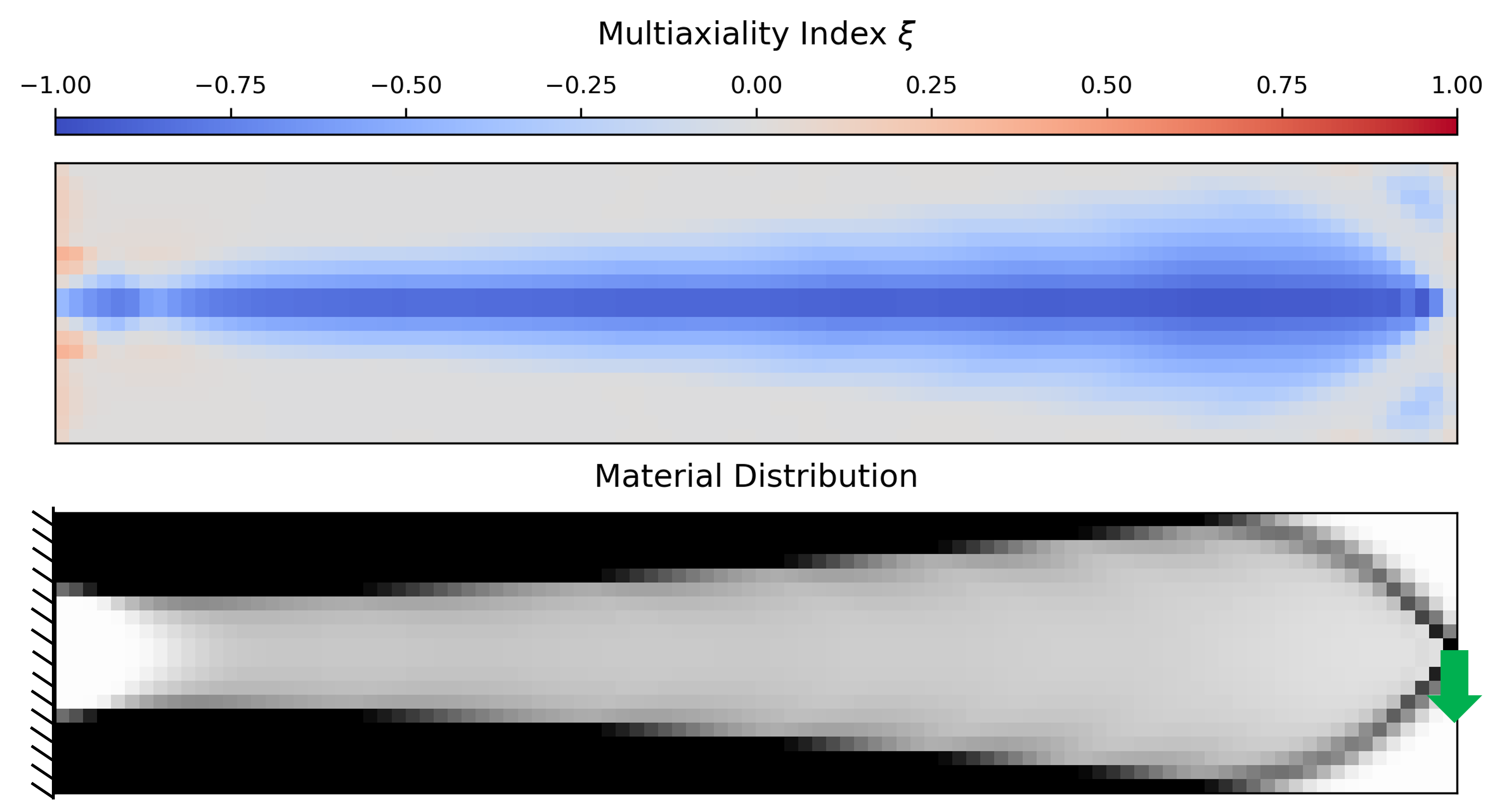}
	\caption{Final topology after iteration 66: continuous gray intermediate densities in the multiaxial shear web (100$\times$20 $Q_4$ elements, no SIMP penalization: $p=1$).}
	\label{beam_p1_l}
	
\end{figure*}

The topology optimization of a cantilever beam provides a representative example in which predominantly uniaxial and multiaxial load-transfer mechanisms coexist within the same structure. 
We first consider again the case without SIMP penalization ($p=1$). The initial stress state and the converged optimized topology are shown in Figures~\ref{beam_p1_0} and~\ref{beam_p1_l}, respectively.

In the initial state, we can observe a pronounced band of elevated multiaxiality along the neutral axis of the beam. This region corresponds to the primary shear-transfer zone connecting the tension and compression chords. As the optimization procedure progresses, the outer chords converge to dense solid regions, whereas the multiaxial shear zone develops into a continuous intermediate-density web.

This result is fully consistent with the interpretation developed so far. 
In regions of multiaxial load transfer, intermediate material distributions are mechanically favorable because they allow stresses to be transmitted simultaneously in multiple directions. In the absence of SIMP penalization, the optimizer therefore naturally converges to continuous gray material within these regions.

\begin{figure*}[t!]
	\centering
	
	\includegraphics[width=0.9\textwidth]{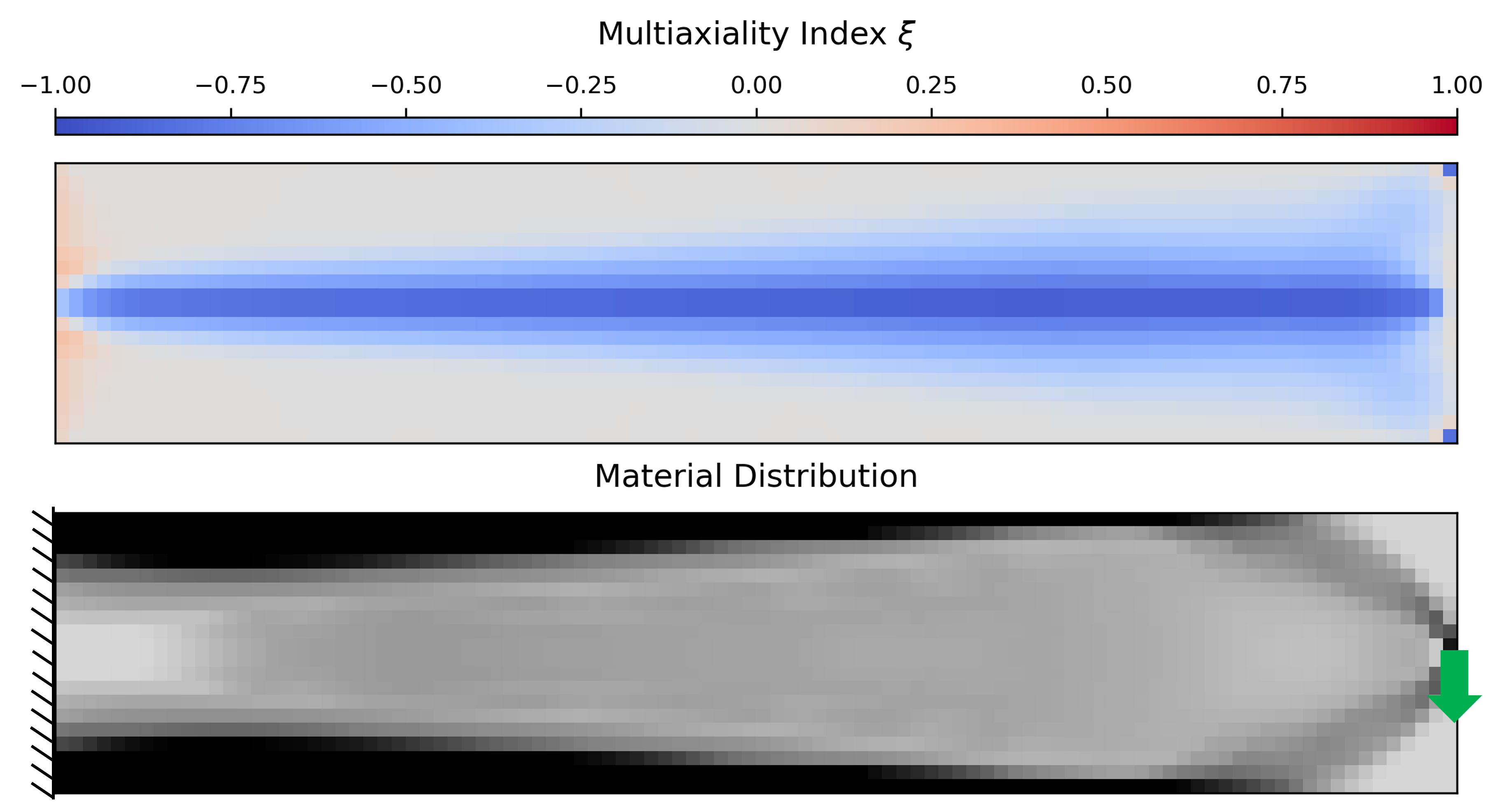}
	\caption{Cantilever beam during the early optimization stage after iteration 5 (100$\times$20 $Q_4$ elements, SIMP penalty factor $p=3$).}
	\label{beam_p3_5}
	
	\vspace{0.2cm} 
	
	\includegraphics[width=0.9\textwidth]{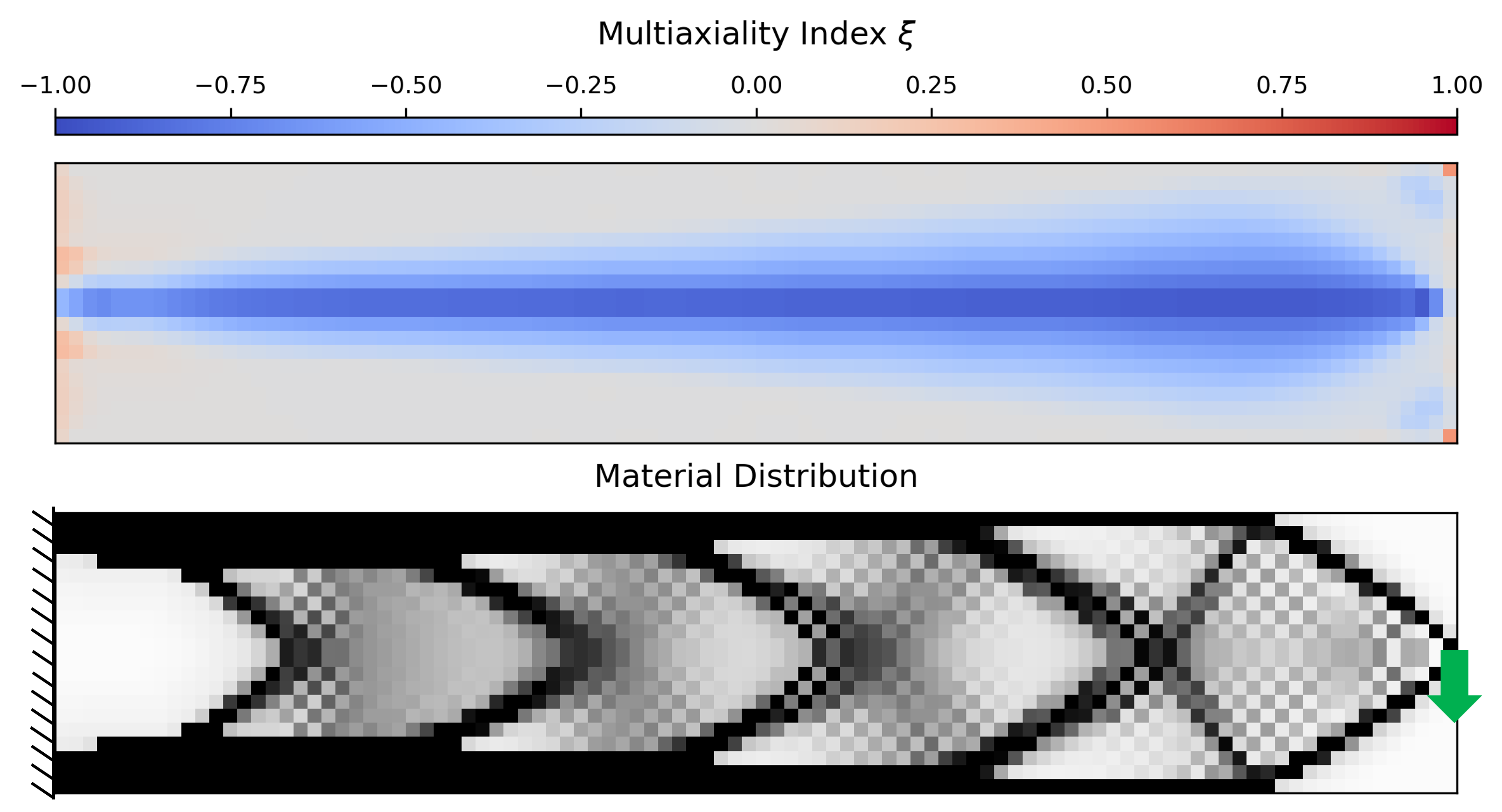}
	\caption{antilever beam during subsequent optimization stage ater iteration 15: growth of checkerboarding within the multiaxial shear regions (100$\times$20 $Q_4$ elements, SIMP penalty factor $p=3$).}
	\label{beam_p3_15}
	
\end{figure*}

The situation changes fundamentally once SIMP penalization is activated. Figures~\ref{beam_p3_5} and~\ref{beam_p3_15} show two representative stages of the optimization process for $p=3$.

The intermediate-density shear web is now replaced by a localized checkerboard pattern, while the outer tension and compression chords remain solid. Importantly, the checkerboard patterns develop precisely within the same multiaxial region that previously contained the gray material in the unpenalized case with $p=1$.

The cantilever benchmark thus provides a direct link between the observations of the previous sections and the locking-based interpretation of \citet{diaz1995checkerboard}. Under uniaxial load transfer, continuous solid struts remain mechanically preferable and checkerboarding does not emerge. Under multiaxial load transfer, however, intermediate material distributions become favorable. Since SIMP penalization suppresses these gray regions, the optimizer exploits the artificial stiffness of the checkerboard patterns as a discrete substitute. Consequently, checkerboarding localizes exactly in those regions where multiaxial load transfer would otherwise require continuous intermediate densities.

\subsection{Necessary and sufficient conditions for checkerboard formation}\label{sec46}

The numerical experiments presented so far have revealed a consistent relationship between checkerboard formation and the underlying stress state. Regions dominated by uniaxial load transfer remain checkerboard-free, whereas checkerboard patterns systematically emerge in regions characterized by multiaxial loading.

However, a closer examination of the benchmark results shows that multiaxiality alone is not sufficient to trigger checkerboard formation. Several counterexamples can be identified. In the compression rod discussed in Section~\ref{sec:tension_rod_hindered}, multiaxial stress states are present both near the constrained support and near the location where the load is applied. Yet, checkerboard patterns develop only along the support at the left boundary. Similarly, the cantilever beam exhibits multiaxial stress states not only within the central shear web, but also in the vicinity of the point load and within the outer corner regions. Again, checkerboard patterns appear only within a subset of these regions.

The distinguishing factor is the structural role of the corresponding domain. Load introduction regions must transfer concentrated forces into the primary load-carrying members of the structure and therefore require dense, continuous material layouts with densities approaching the upper bound. Since checkerboard patterns contain approximately 50\,\% void space, they are mechanically ill-suited for this task. Conversely, corner regions and other zero-force domains require little or no material at all, causing the density to approach its lower bound. Although multiaxial stress states may formally be present in these areas, no load-bearing material is required and checkerboards therefore cannot develop.

These observations indicate that checkerboard formation requires not only multiaxial loading but also a structural demand for intermediate load-bearing capacity. In terms of classical strut-and-tie models, these conditions are naturally fulfilled within structural shear webs, whereas tension and compression chords correspond to predominantly uniaxial load transfer and therefore remain checkerboard-free. 

In multiaxial load-transfer zones, stresses must be transmitted simultaneously in multiple directions. In the absence of SIMP penalization, these regions evolve towards continuous intermediate-density distributions, as demonstrated by the cantilever beam in Section~\ref{sec:cantilever_beam}. Once SIMP penalization is activated, however, these gray regions become energetically unfavorable. The optimizer therefore seeks alternative configurations that preserve their mechanical functionality. 

For bilinear $Q_4$ elements, the checkerboard pattern provides precisely such an alternative. Owing to its artificially high stiffness, it acts as a discrete surrogate for the suppressed gray material. Consequently, checkerboard patterns emerge only when the following two conditions are simultaneously fulfilled:

\begin{enumerate}
\item The local stress state must be sufficiently multiaxial.
\item The corresponding region must require intermediate load-bearing capacity, i.e., neither fully solid nor fully void material distributions are mechanically optimal.
\end{enumerate}

Multiaxiality therefore constitutes a necessary but not sufficient condition for checkerboard formation. The actual localization of checkerboard patterns is governed by the interplay between multiaxial load transfer, the structural demand for intermediate material, and the artificial stiffness introduced by the locking behavior of bilinear $Q_4$ elements.

\subsection{Coordinate invariance and mesh orientation}

A potential objection to the preceding interpretation is that checkerboard patterns may simply be a consequence of the alignment between the stress field and the finite element mesh. To assess this possibility, we repeat the cantilever benchmark with a rotated design domain. Specifically, the load application and boundary conditions are rotated by $45^\circ$ while the underlying finite element discretization remains unchanged.

\begin{figure}[h!]
	\centering
	\includegraphics[width=0.95\columnwidth]{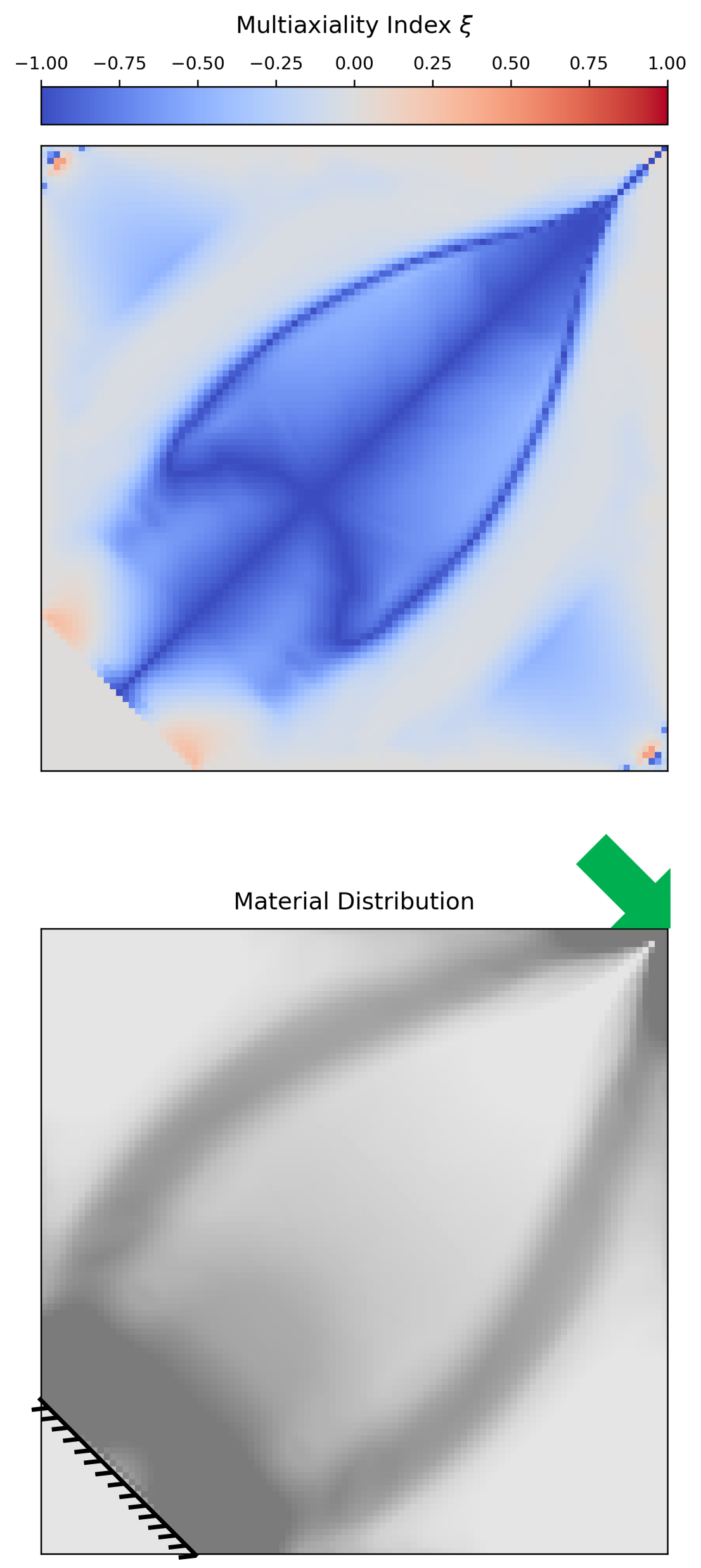}
	\caption{Cantilever beam with rotated design domain: early stage after iteration 4 (100$\times$100 $Q_4$ elements, SIMP penalty factor $p=3$).}
	\label{Rbeam_4}
\end{figure}
\begin{figure}[h!]
    \centering
	\includegraphics[width=0.95\columnwidth]{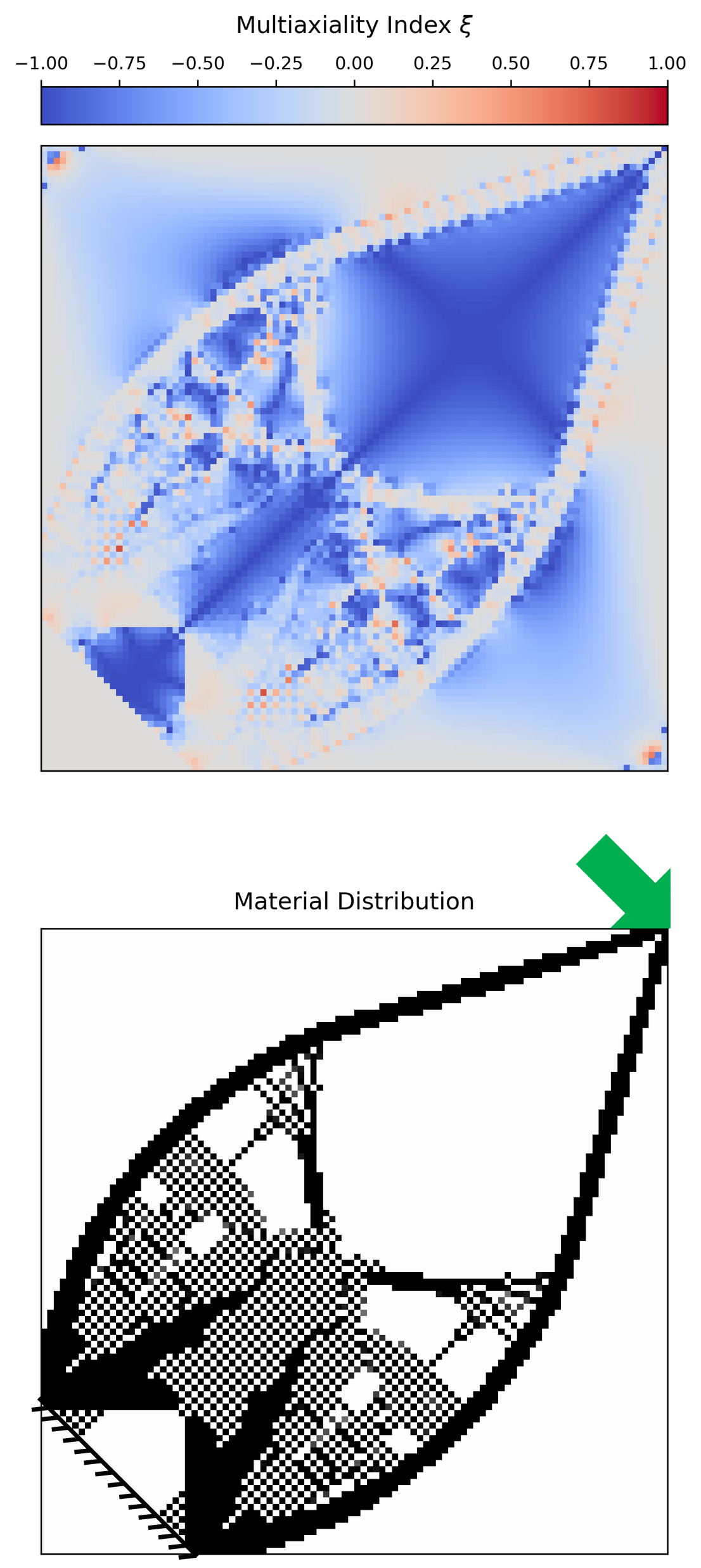}
	\caption{Cantilever beam with rotated design domain: final topology after 99 iterations (100$\times$100 $Q_4$ elements, SIMP penalty factor $p=3$).}
	\label{Rbeam_99}
\end{figure}

Figures~\ref{Rbeam_4} and~\ref{Rbeam_99} show the distribution of the multiaxiality index $\xi$ and the material distribution during the early stage of the iterative optimization procedure and for the final optimized result, respectively. While the rotated benchmark produces a different optimal structural layout, the characteristic localization behavior of the checkerboard patterns remains unchanged. Checkerboards again emerge within regions of elevated multiaxiality, whereas regions of predominantly uniaxial load transfer remain checkerboard-free. 

This observation is fully consistent with the necessary and sufficient
conditions for checkerboard formation proposed in Section~\ref{sec46}. The multiaxiality index $\xi$ derived from the principal stresses represents an invariant measure of the local stress state. Consequently, the structural conditions promoting checkerboard formation are independent of the chosen coordinate system. The persistence of the localization pattern under rotation therefore provides additional evidence that checkerboard formation is governed by the underlying stress state rather than by a particular alignment of the mesh.

\begin{figure*}[t!]
	\centering
	
	\includegraphics[width=0.9\textwidth]{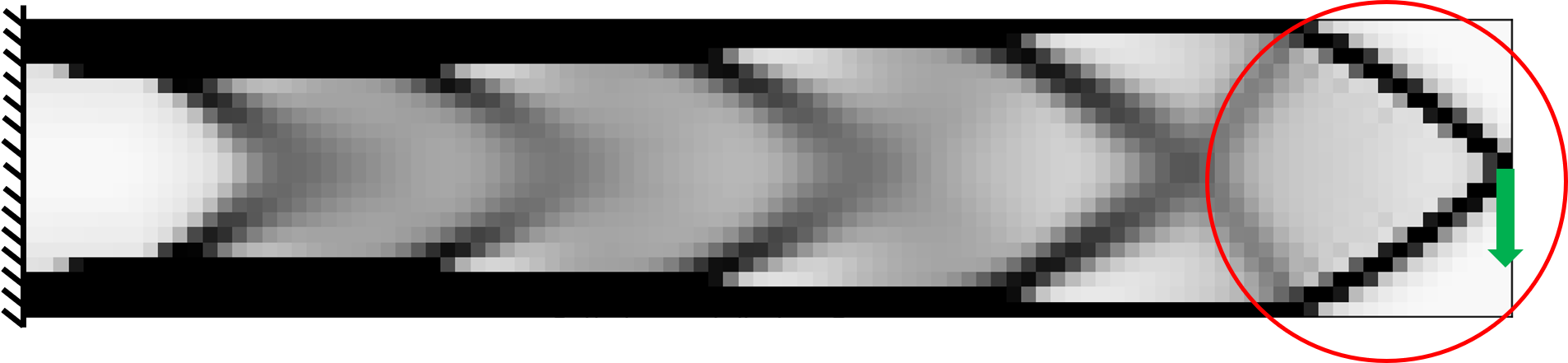}
	\caption{Material distribution showing early checkerboard seeding with $Q_9$ elements.}
	\label{q9_seeding}
	
	\vspace{0.2cm} 
	
	\includegraphics[width=0.9\textwidth]{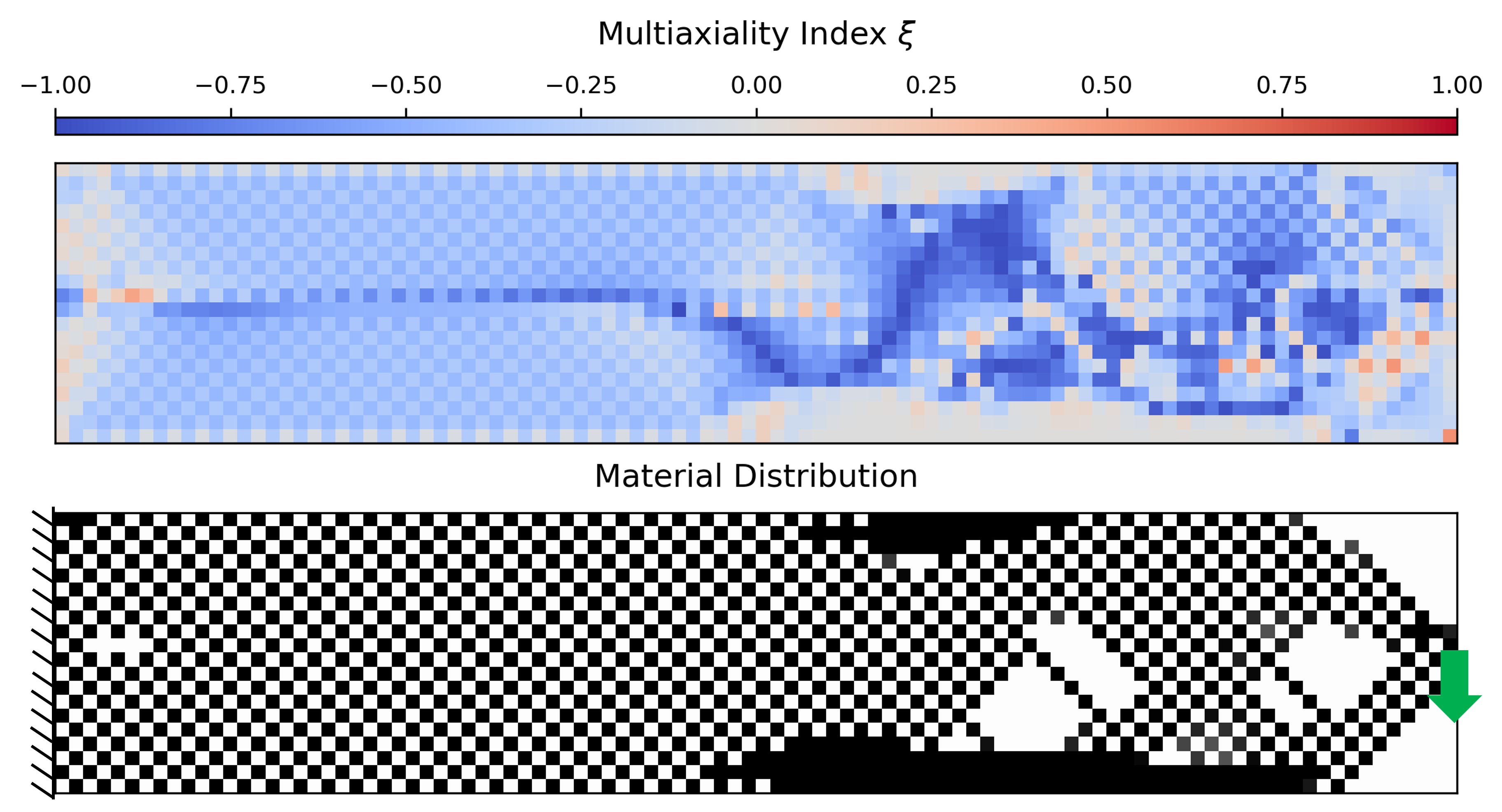}
	\caption{Converged topology of a cantilever beam, discretized with $Q_9$ elements and initialized with a full checkerboard pattern (100$\times$20 $Q_9$ elements, SIMP penalty factor $p=3$, 79 iterations).}
	\label{q9_checker}
	
\end{figure*}

\subsection{Seeding, persistence, and local minima}\label{seeds}

The previous sections established that checkerboard formation is triggered by multiaxial load transfer and promoted by the artificial stiffness of bilinear finite elements. However, the emergence of a checkerboard is not instantaneous. In practice, the pattern is initiated by small numerical perturbations in the sensitivities of neighboring elements, a process referred to here as \emph{seeding}.

This distinction is important because the absence of checkerboarding in a discretization may arise either from preventing the initial amplification of checkerboard seeds or from eliminating already-established checkerboard configurations.

To investigate the role of seeding independently of locking, the cantilever benchmark is analyzed using biquadratic $Q_9$ elements. Figure~\ref{q9_seeding} shows an early optimization stage. Although localized checkerboard seeds can still be observed, the pattern is not amplified during the subsequent optimization process. Instead, the optimizer progressively removes these perturbations and converges towards a smooth material distribution.

This behavior confirms that checkerboard initiation and checkerboard growth are fundamentally different phenomena. Numerical perturbations may generate local checkerboard seeds even in stable discretizations. However, in contrast to bilinear $Q_4$ elements, biquadratic $Q_9$ elements do not exhibit the artificial stiffness overestimation associated with checkerboard patterns and therefore provide no incentive for their further amplification.

Interestingly, once a fully developed checkerboard pattern is established, it can persist even for $Q_9$ elements. To demonstrate this effect, the cantilever beam is initialized directly as a complete checkerboard pattern and subsequently optimized using $Q_9$ elements. The resulting topology is shown in Figure~\ref{q9_checker}. Despite the absence of locking-induced stiffness overestimation, the optimization process retains significant portions of the initially prescribed checkerboard pattern.

This observation highlights the strongly non-convex nature of the optimization problem. While $Q_9$ elements suppress the growth of checkerboard seeds, they do not necessarily eliminate already-established checkerboard configurations. The results suggest that a fully developed checkerboard may constitute a local optimum from which the gradient-based optimization algorithm cannot easily escape. Consequently, the absence of checkerboarding in practice depends not only on the mechanical stability of the discretization, but also on the evolutionary path followed during the optimization process.

The results therefore suggest a clear distinction between three mechanisms: seeding, amplification, and persistence. Seeding may occur in both $Q_4$ and $Q_9$ discretizations. Amplification is characteristic of locking-prone $Q_4$ elements and is responsible for the formation of pronounced checkerboard patterns. Persistence, on the other hand, appears to be associated with local minima and may therefore occur even in discretizations that do not promote checkerboard growth.

\section{Summary, conclusions and outlook}\label{fazit}

Checkerboard patterns are among the most prominent numerical artifacts in density-based topology optimization using the SIMP method and linear displacement-based finite elements. Historically, this phenomenon has primarily been attributed either to a violation of the LBB condition in the context of mixed finite element formulations \citep{jog1996stability} or to the artificial stiffness overestimation of periodic checkerboard microstructures \citep{diaz1995checkerboard}.

In this work, we demonstrated that neither perspective provides a complete explanation in isolation. Since classical density-based optimization algorithms employ a decoupled update scheme and strict bounding-box constraints, the LBB condition does not directly apply. Conversely, while the micromechanical analysis of \citet{diaz1995checkerboard} successfully explains the local stiffness overestimation of checkerboard patterns, it does not directly explain their macroscopic localization in specific regions of the design domain. We showed that the missing link is provided by the structural load-transfer mechanism governing the global stress field.

Through systematic numerical investigations, we identified multiaxial load transfer as the fundamental driver of checkerboard localization. Regions dominated by uniaxial tension or compression consistently converged to solid, checkerboard-free configurations, whereas checkerboards systematically emerged in regions characterized by multiaxial stress states. However, multiaxiality alone was found to be insufficient. Checkerboard formation requires two conditions to be simultaneously fulfilled: a sufficiently multiaxial stress state and a structural demand for intermediate load-bearing capacity.

From a structural mechanics perspective, regions of multiaxial load transfer naturally evolve towards continuous intermediate-density material distributions. 
Once SIMP penalization is activated, however, these gray regions become energetically unfavorable. For bilinear $Q_4$ elements, the checkerboard pattern provides a discrete surrogate for this suppressed intermediate material. Owing to the locking-induced stiffness overestimation identified by \citet{diaz1995checkerboard}, the optimizer can exploit checkerboard patterns to mimic the mechanical functionality of unpenalized gray material while avoiding the SIMP penalty.

The numerical experiments further demonstrated that asymptotic stiffness equivalence alone is insufficient to explain checkerboard formation. While infinitely periodic checkerboards are asymptotically as stiff as continuous struts under uniaxial loading, finite checkerboard patches exhibit pronounced boundary effects that render them mechanically inferior. 

Finally, the distinction between seeding, amplification, and persistence was investigated. Numerical experiments with biquadratic $Q_9$ elements demonstrated that checkerboard seeds may arise even in discretizations that do not exhibit locking-induced stiffness overestimation. However, in contrast to bilinear $Q_4$ elements, these perturbations are not amplified during the optimization process. At the same time, our results indicate that fully developed checkerboard configurations may persist as robust local optima, highlighting the strongly non-convex nature of the optimization problem.

The mechanical interpretation developed in this work also provides a new perspective on regularization methods. Since checkerboard formation originates from the inability of linear elements to correctly represent the mechanical weakness of neighboring checkerboard cells, successful regularization techniques may be viewed as a means of introducing additional information exchange between adjacent elements. From this perspective, sensitivity filtering and gradient-based approaches can be interpreted as restoring a form of nonlocal mechanical awareness that suppresses the locking-driven stiffness overestimation. These observations may provide useful guidance for the development of future regularization strategies that address checkerboarding without unnecessarily diffusing sharp structural boundaries.

In summary, the results establish checkerboard formation and localization as the consequence of a three-way interaction between multiaxial load transfer, SIMP penalization, and locking-induced stiffness overestimation. Checkerboards emerge precisely in those regions where multiaxial load transfer creates a structural demand for intermediate densities that are subsequently suppressed by SIMP penalization.

\section*{Statements and declarations}

\textbf{Funding.} Open access funding enabled and organized by Projekt
DEAL.\\
\newline
\noindent\textbf{Author contributions.} \textbf{Iulian Paunel:} Conceptualization, Methodology, Software, Formal analysis, Investigation, Writing -- Original Draft, Visualization. \textbf{Jonathan Stollberg:} Conceptualization, Methodology, Writing -- Review \& Editing, Supervision. \textbf{Dominik Schillinger:} Conceptualization, Writing -- Review \& Editing, Supervision, Project administration, Funding acquisition.\\
\newline
\noindent\textbf{Conflict of interest.} The authors declare that they have no known competing interests or personal relationships that could have appeared to influence the work reported in this paper.\\
\newline
\noindent\textbf{Replication of results.} The results can be reproduced based on the information presented in the manuscript only. All data and the developed Python code can be made available upon request from the corresponding author.

\bibliography{bibliography}

\end{document}